\documentclass[
 amssymb, amsmath,
 reprint,%
aps, prl,
]{revtex4-1}
\usepackage{xcolor}

\usepackage{rotating}
\usepackage{graphicx}
\usepackage{scrextend}
\usepackage{simplewick}

\usepackage{chemmacros}
\chemsetup{
 formula = mhchem,
 greek = chemgreek,
 modules = thermodynamics,
 modules = redox,
 modules = reactions
}

\usepackage{bm}
\usepackage[colorlinks=true,linkcolor=blue]{hyperref}
\expandafter\ifx\csname package@font\endcsname\relax\else
 \expandafter\expandafter
 \expandafter\usepackage
 \expandafter\expandafter
 \expandafter{\csname package@font\endcsname}
\fi
\begin{document}

\title{Predictive simulations of ionization energies of solvated halide ions with relativistic embedded Equation of Motion Coupled-Cluster Theory}

\author{Yassine Bouchafra}
\affiliation{Universit\'e de Lille, CNRS, UMR 8523 -- PhLAM -- Physique des Lasers, Atomes et Mol\'ecules, F-59000 Lille, France Tel: +33-3-2043-4163}

\author{Avijit Shee}
\affiliation{Department of Chemistry, University of Michigan, 930 N.\ University, Ann Arbor, MI 48109-1055, USA} 
\altaffiliation{Universit\'e de Lille, CNRS, UMR 8523 -- PhLAM -- Physique des Lasers, Atomes et Mol\'ecules, F-59000 Lille, France Tel: +33-3-2043-4163}

\author {Florent R\'eal}
\affiliation{Universit\'e de Lille, CNRS, UMR 8523 -- PhLAM -- Physique des Lasers, Atomes et Mol\'ecules, F-59000 Lille, France; Tel: +33-3-2043-4163}
\author {Val\'erie Vallet}
\affiliation{Universit\'e de Lille, CNRS, UMR 8523 -- PhLAM -- Physique des Lasers, Atomes et Mol\'ecules, F-59000 Lille, France; Tel: +33-3-2043-4163}
\author{Andr\'e Severo Pereira Gomes}
\email{andre.gomes@univ-lille.fr (corresponding author)}
\affiliation{Universit\'e de Lille, CNRS, UMR 8523 -- PhLAM -- Physique des Lasers, Atomes et Mol\'ecules, F-59000 Lille, France; Tel: +33-3-2043-4163}

\date{\today}
\revised{\today}

\begin{abstract}
A subsystem approach for obtaining electron binding energies in the valence region and apply it to the case of halide  ions (\ce{X-}, X = F--At) in water is presented. This approach is based on  
electronic structure calculations combining the relativistic equation of motion coupled-cluster method for electron detachment (EOM-IP-CCSD) and density functional theory via the frozen density embedding (FDE) approach, using structures from classical molecular dynamics with polarizable force fields for discrete systems (in the present study, droplets containing the anion and 50 water molecules).
Our results indicate one can accurately capture both the large solvent effect observed for the halides as well as the splitting of their ionization signals due to the increasingly large spin-orbit coupling of the p$_{3/2}$--p$_{1/2}$ manifold across the series, at an affordable computational cost. Furthermore, due to the quantum mechanical treatment of both solute and solvent, electron binding energies of semi-quantitative quality are also obtained for (bulk) water as by-products of the calculations for the halogens (in droplets). 
\end{abstract}

\maketitle

Photoelectron (PE) spectroscopy~\cite{Bahr:1973hw} is a particularly powerful technique (nowadays often complemented by electronic structure calculations) to investigate bound states at the valence or inner regions, either to obtain information on the nature of bonding for species in the gas-phase~\cite{Dau:2012fua,Li:2014cm,Su:2015hv}, in solution~\cite{Seidel:2016iga,Pham:2017dy} or at interfaces~\cite{Bagus:2013jq,Trotochaud:2016ip,KnopGericke:2017hv} as well as to follow and identify chemical changes in complex media~\cite{Kong:2017hs,BartelsRausch:2017bv,Raheem:2017jg}.
Such techniques have been extensively used to investigate species such as halogens and halogen-containing species~\cite{Tsal:1975era,Gilles:1992cc,Lago:2005kr}, which are of great importance in atmospheric processes~\cite{Simpson:2015bz,SaizLopez:2011is} such as photochemical reactions leading to ozone depletion, or aerosol formation~\cite{GomezMartin:2013faa}. 

The simplest halogenated systems of relevance are the halides, originating mostly from marine aerosols~\cite{Carpenter:2015hy}, and understanding how these species interact with water is, apart from its intrinsic interest, of importance for better understanding their effects in the environment.
Experimental studies on clusters~\cite{Markovich:1994he} and bulk~\cite{Winter:2005kx} aqueous solutions have established that there are very large shifts in the PE spectrum of the halides upon solvation, highlighting strong interactions between the anions and the water solvent. 
Early theoretical studies determined the halides' electron binding energies (BEs) employing ab initio calculations~\cite{Pathak:2008bc,Dolgounitcheva:2012ik,He:2018en} or combining these with classical molecular dynamics simulations with periodic boundary conditions~\cite{Winter:2005kx}. These studies indicate that not including specific interactions (hydrogen bond etc.) between the halogens and the solvent water molecules leads to a poor description of the halide BEs~\cite{Winter:2005kx,Coons:2018ba}, apart from the fact that quantum-classical approaches cannot yield the electronic structure of the solvent.

Currently the most sophisticated theoretical approaches to obtain PE spectra for the whole system quantum-mechanically (``full-QM'') rely upon density functional theory (DFT) to obtain the ground-state for the solvent-solute system (as in Car-Parrinello molecular dynamics (CPMD)~\cite{Hutter:2011gl}) followed by use of many-body Green's function (MBGF)-based perturbation theories (e.g.\ $GW$ and variants such as $G_0W_0$~\cite{Zhang:2013bv,Gaiduk:2016cf,Gaiduk:2017bq,Pham:2017dy,Gaiduk:2018dd,Gaiduk:2018hm}). MBGF approaches are not without downsides, however: the first is their high computational cost for fully self-consistent variants, especially if the calculations employ periodic boundary conditions and require large (super)cells. A second, and more serious issue is the lack of exchange diagrams in self-energy beyond first order. This is particularly  a shortcoming in the treatment of molecular systems. 

$GW$-based approaches have been shown to introduce relatively large errors for the calculation of BEs~\cite{Blase:2016hr,*Blase:2016hr2,Lange:2018di}, compared to reference single-reference coupled-cluster (CCSD(T)) or equation-of-motion coupled cluster (EOM-IP-CCSD)~\cite{Bartlett:2007kv,Bartlett:2011hoa} calculations. Recent benchmarking studies suggest even lower-scaling, approximate variants to EOM-CCSD~\cite{Goings:2014dt,Dutta:2018hi}, can be competitive in accuracy with $GW$ calculations of ionizations and electron affinities, and especially so for $G_0W_0$~\cite{Lange:2018di}.

This communication presents a full-QM electronic structure approach for obtaining BEs of discrete systems such as water-halide ion (\ce{X-}, X = F--At) aggregates, as a cost-effective yet accurate alternative to $GW$-based calculations, by coupling relativistic EOM-IP-CCSD calculations for the halides (since relativistic effects, and in particular spin-orbit coupling (SOC)~\cite{Saue:2011ir}, on the BEs are increasingly important along the halogen series) and scalar relativistic DFT calculations for the water molecules through the frozen density embedding (FDE) method~\cite{Cortona:1991fm,Cortona:1992br,Wesolowski:1993cv}. 

The key idea of FDE (see~\cite{SeveroPereiraGomes:2012dv,Jacob:2014eb,Wesolowski:2015kz,Sun:2016hy} for further details and its relationship to other embedding methods) is the partitioning of a system's electron density $n(\boldsymbol{r})$ into a number of fragments (for simplicity here two such fragments are considered, so $n(\boldsymbol{r}) =n_\text{I}(\boldsymbol{r}) + n_\text{II}(\boldsymbol{r})$) and total energy $E[n(\boldsymbol{r})]$, which can be rewritten as a sum of subsystem energies ($E_{i}[n_{i}(\boldsymbol{r})], i=\text{I,II}$) plus an interaction energy ($E_{(\text{int})}$)
\begin{equation}
E[n] = E_\text{I}[n_\text{I}] + E_\text{II}[n_\text{II}] + E_{(\text{int})}[n_\text{I}, n_\text{II}].
\label{eq:Etot}
\end{equation}
The latter collects the inter-subsystem interaction terms, 
\begin{align}
E_{(\text{int})}[n_\text{I}, n_\text{II}] 
    &=  \int \left[ n_\text{I}(\boldsymbol{r}) v_\text{nuc}^\text{II}(\boldsymbol{r})  +  n_\text{II}(\boldsymbol{r}) v_\text{nuc}^\text{I}(\boldsymbol{r}) \right]d\boldsymbol{r} \nonumber \\
    & + \int \int \frac{n_\text{I}(\boldsymbol{r}) n_\text{II}(\boldsymbol{r}')}{|\boldsymbol{r} - \boldsymbol{r}'|}d\boldsymbol{r}d\boldsymbol{r}'  \nonumber\\
     & + E_\text{xck}^\text{nadd}[n_\text{I}, n_\text{II}] + E_\text{nuc}^\text{I,II},
         \label{eq:Eint}
\end{align}
where $v_\text{nuc}^{i}$ is the nuclear potential ($i=\text{I,II}$), $E_\text{nuc}^\text{I,II}$ the nuclear repulsion energy between subsystems and $E_\text{xck}^\text{nadd}$ accounts for non-additive contributions due to exchange-correlation (xc) and kinetic energy (k) contribution. $E_\text{xck}^\text{nadd}$ is defined as
\begin{align}
E_\text{xck}^\text{nadd}[n^\text{I}, n^\text{II}] 
&= E_\text{xc}^\text{nadd}[n^\text{I}, n^\text{II}] + T_\text{s}^\text{nadd}[n^\text{I}, n^\text{II}] \nonumber \\
&= E_\text{xc}[n^\text{I} + n^\text{II}] - E_\text{xc}[n^\text{I}] - E_\text{xc}[n^\text{II}] \nonumber \\
&+ T_\text{s}[n^\text{I} + n^\text{II}] - T_\text{s}[n^\text{I}] - T_\text{s}[n^\text{II}].
\end{align}
The non-additive kinetic energy contribution provides a repulsive interaction that offsets the attractive interaction between the nuclear framework of one subsystem and the density of the other~\cite{Roncero:2008do} which, if not properly matched, can lead to spurious delocalization of the electron density of one subsystem over the region of other~\cite{Jacob:2007hs} (as seen, for instance, in point-charge or QM/MM embedding~\cite{Reinholdt:2017gl}). For reasons of computational efficiency, the FDE calculations in this work employ approximate kinetic energy density functionals~\cite{Lembarki:1994fx} which provide good but nevertheless limited accuracy~\cite{Gotz:2009cw} for systems such as those discussed here, which are not covalently bound.

In a purely DFT framework, the density for a subsystem of interest $n_\text{I}$ is obtained by minimizing the total energy (Eq.~\ref{eq:Etot})  with respect to variations on $n_\text{I}$ while keeping $n_\text{II}$ frozen, yielding Kohn-Sham-like equations
\begin{equation}
\left[ T_\text{s}(i) + v_\text{KS}[n_\text{I}] + v_\text{int}^\text{I}[n_\text{I}, n_\text{II}] - \varepsilon_i 
\right] \phi_i^\text{I}(\boldsymbol{r})
= 0,
\label{eq:KSCED}
\end{equation}
where $v_\text{KS}[n_\text{I}]$ and $T_\text{s}(i)$ are the usual Kohn-Sham potential and kinetic energy (from ${\delta E_\text{I}[n_\text{I}]}/{\delta n_\text{I}}$), and
\begin{align}
v_\text{int}^\text{I}(\boldsymbol{r}) &= v_\text{xc}^\text{nadd}(\boldsymbol{r}) + \left. \frac{\delta T_\text{s}^\text{nadd}}{\delta n} \right|_{n_\text{I}} + v_\text{nuc}^\text{II}(\boldsymbol{r})\nonumber\\
&+ \int \frac{n_\text{II}(\boldsymbol{r}')}{|\boldsymbol{r} - \boldsymbol{r}'|}d\boldsymbol{r}'
\label{embpot-formula}
\end{align}
is the embedding potential (from ${\delta E_{(\text{int})}[n_\text{I}, n_\text{II}]}/{\delta n_\text{I}}$), which describes the interaction between subsystems. 

FDE provides a formally exact framework that allows DFT to be replaced by wavefuction theory (WFT)-based treatments for one~\cite{Govind:1999ff,Wesoiowski:2008fh,Gomes:2008bz,Hofener:2012cr} (WFT-in-DFT) or all subsystems~\cite{Hofener:2012bv} (WFT-in-WFT), with the embedding potential being calculated from Eq.~\ref{embpot-formula} irrespective of the level of electronic structure employed, though using the electron densities from the respective methods.
Obtaining electron densities for WFT methods in general and coupled-cluster in particular is computationally expensive (the latter requiring the solution of the CC $\Lambda$-equations ground state~\cite{Bartlett:2007kv}), and it has been found that an approximate scheme--where $v_\text{int}^\text{I}$ is obtained from preparatory DFT-in-DFT calculations~\cite{Gomes:2008bz,Hofener:2013eia} and treated as a (local) one-electron operator added to the Fock matrix in the WFT calculations--works very well in practice. This latter prescription is the one followed here. 

In the EOM-IP-CCSD method, BEs are obtained from the solution of the eigenvalue equation~\cite{Bartlett:2011hoa,shee:arxiv:2018}
\begin{equation}
(\overline{H} R^\text{IP}_k)_c = \Delta E_k R^\text{IP}_k  \label {Eq:energy}
\end{equation}
where $\Delta E_k$ is the $k$-th ionization energy for the system, $\overline{H} = e^{-T}\hat{H}e^T$ is the (CCSD) similarity transformed Hamiltonian (here including $v_\text{int}^\text{I}(\boldsymbol{r})$) and 
\begin{equation}
R^\text{IP} = \sum_i r_i \{i\} + \sum_{i>j,a} r_{ij}^{ a} \{a^{\dag} j i\} 
\end{equation}
the wave operator that transforms the CC ground-state to the electron detachment states.

In the preparatory DFT-in-DFT calculations the SAOP model potential~\cite{Gritsenko:1999bt} has been used. This potential is constructed to yield Kohn-Sham potentials showing proper atomic shell structure and correct asymptotic behavior, and with it calculations have a computational cost equivalent to Kohn-Sham DFT using GGAs. The SAOP orbital energies have been shown to provide BEs that in very good agreement with coupled cluster calculations~\cite{Tecmer:2011cf}. Given the evidence in the literature that Kohn-Sham densities obtained with functionals yielding accurate BEs compare quite well to densities obtained with coupled cluster methods~\cite{Grabowski:2011ct,Ranasinghe:2017ec}, a $v_\text{int}^\text{I}$ obtained with SAOP densities should provide a good approximation to one obtained with coupled cluster densities, with the advantage that one obtains a representation for the PE spectrum of water at no additional cost. 

\begin{figure}[ht]
\centering
\begin{minipage}[b]{.32\linewidth}
\centering
\includegraphics[width=\textwidth]{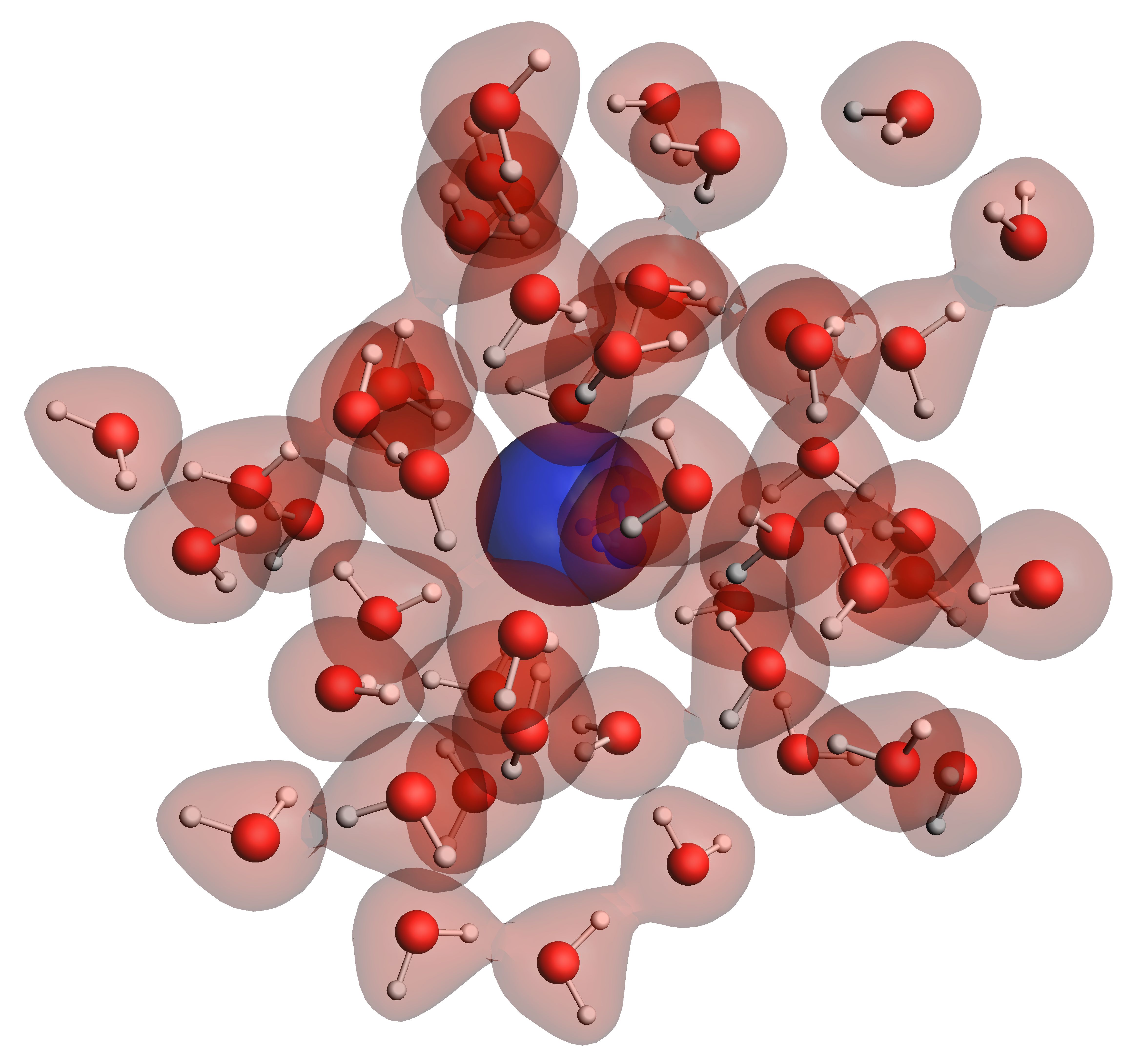}
\end{minipage}
\begin{minipage}[b]{.32\linewidth}
\centering
\includegraphics[width=\textwidth]{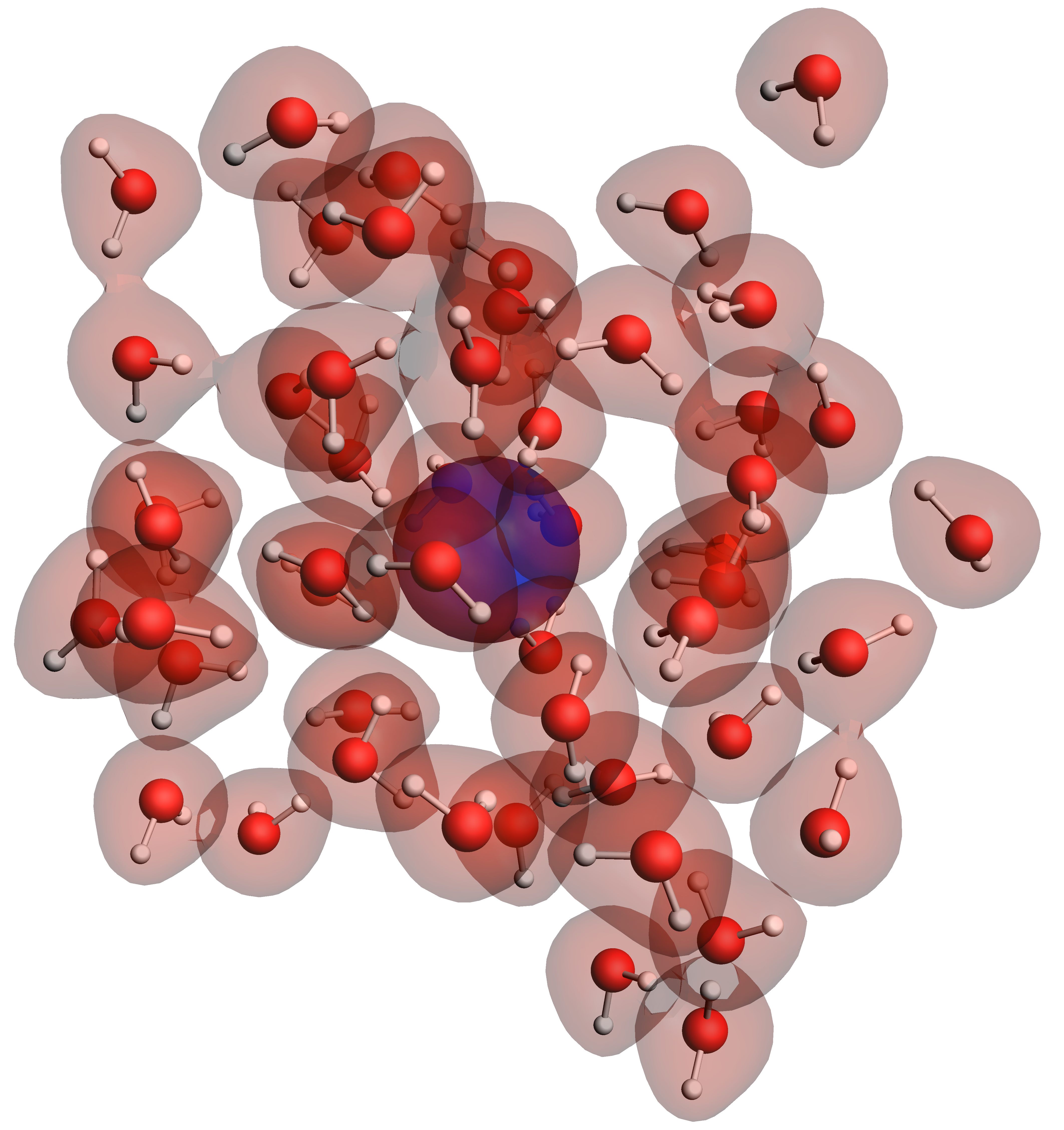}
\end{minipage}
\begin{minipage}[b]{.32\linewidth}
\centering
\includegraphics[width=\textwidth]{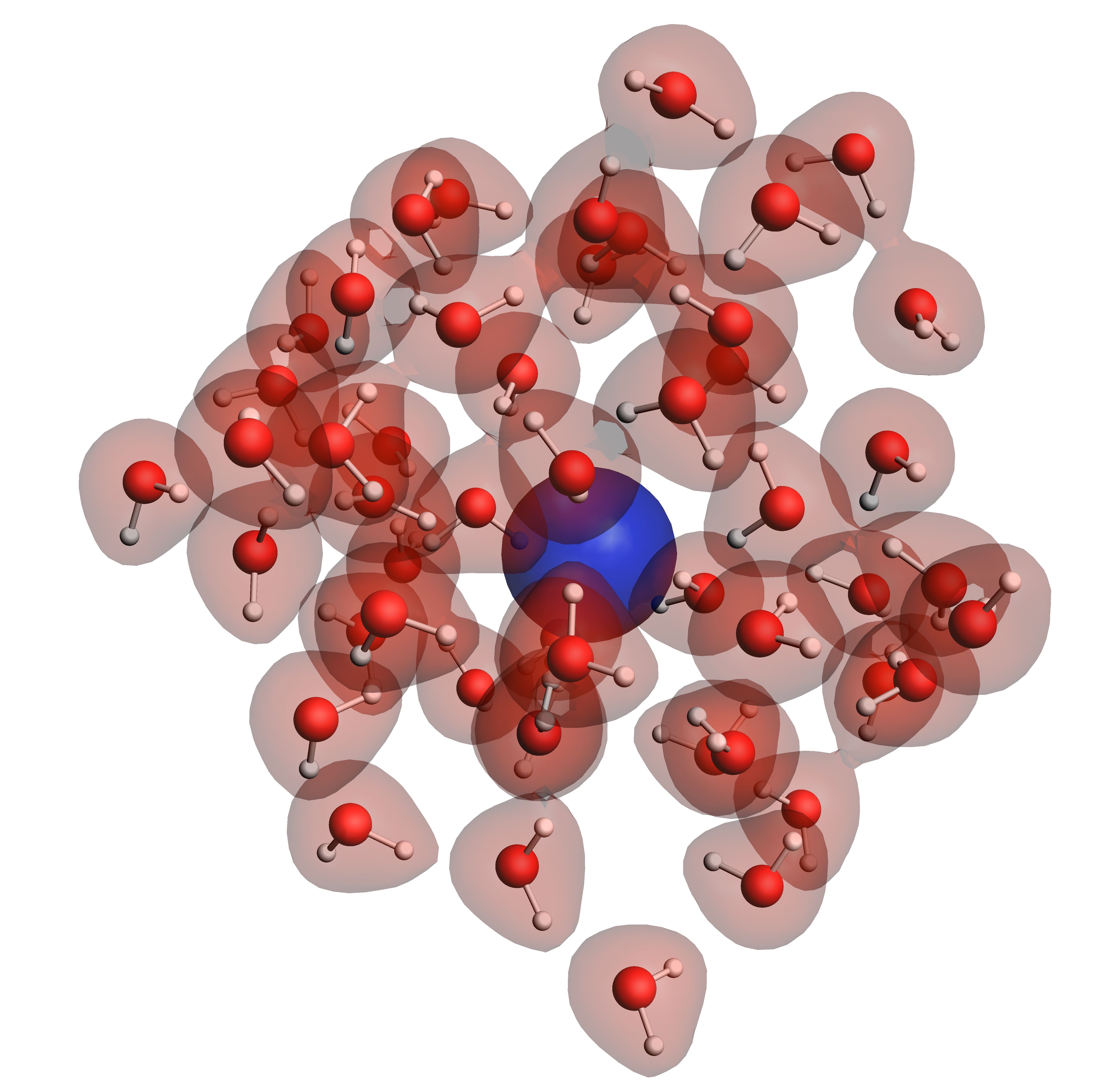}
\end{minipage}
\caption{Views along the (x,y,z) axes for a sample configuration of the CMD simulation for \ce{I-}. The (frozen) density for the water subsystem ($n_\text{II}$) is  superimposed onto the structures~\cite{paper:figures}.}
\label{model}
\end{figure}

The FDE calculations were performed on structures obtained with classical molecular dynamics simulations (CMD) on water-halide droplets containing 50 water molecules and constraining the halogen to be fixed at the droplet's center of mass, using the POLARIS(MD) code~\cite{Masella:2006da,Masella:2011dy,Masella:2013iw,Coles:2015er} and many-body force-fields~\cite{Real:2016bhb} accounting for both polarization effects and the bonding effects within the water molecules (hydrogen bonds) and between the halide and first-hydration shell water units (strong hydrogen bond). From these, after equilibration of the system, were extracted 200 snapshots, which were verified uncorrelated for the BEs (see supplementary information). A particular feature of the droplet structures for all halogen species, such as shown in Fig.~\ref{model} for a snapshot of solvated \ce{I-}, is that the water distribution around the anion is not spherical but elongated, due to strong polarization effects that favor disymmetrised structures, with about six to eight water molecules making up the first solvation shell.

The total system was partitioned into two subsystems, the halide (subsystem I) and the 50 water molecules (subsystem II), corresponding to the simplest partition to calculate the halide BEs (referred to as [\ce{X-}@\ce{(H2O)_{50}}]). This choice is supported by benchmark tests (see supplementary information) as well as prior calculations on small halide-water clusters~\cite{Dolgounitcheva:2012ik}, which show that for \ce{Cl-} the valence ionizations are mostly coming from the halide. For \ce{F-} on the other hand, there are important contributions from both the halogen and the waters (with ionization from the latter being lower in energy than from the halide), and because of this a second model was considered, in which the nearest eight water molecules are also included in subsystem I (referred to as [\ce{(F(H2O)8)-}@\ce{(H2O)_{42}}]). 

The DFT-in-DFT $v_\text{int}$ were obtained over 200 CMD snapshots with the \textsc{PyADF} scripting environment~\cite{Jacob:2011ew}, which used the subsystem DFT implementation in the ADF code~\cite{Jacob:2008fa}, and employed the scalar relativistic (SR) zero-order regular approximation (ZORA) Hamiltonian~\cite{Lenthe:1993ia} and  triple-zeta (TZ2P) quality basis sets~\cite{vanLenthe:2003hq} for all atoms. The non-additive kinetic energy and exchange-correlation contributions to $v_\text{int}$  were calculated with the PW91k~\cite{Lembarki:1994fx} and PBE~\cite{Perdew:1996iq,*Perdew:1996iq2} density functionals, respectively. Unless otherwise noted, all SR-ZORA DFT-in-DFT calculations reported use the same computational setup. 
The embedded EOM-IP-CCSD (EOM) calculations were performed over a subset of 100 CMD snapshots from the originally selected 200 snapshots (see supplementary information) with a  development  version (revisions \texttt{e25ea49} and \texttt{7c8174a})~\cite{shee:arxiv:2018} of the \textsc{Dirac} electronic structure code~\cite{DIRAC17}, using the Dirac-Coulomb (DC) Hamiltonian~\cite{Saue:2011ir,Visscher:1997gh} and uncontracted augmented triple-zeta quality~\cite{Dyall:2002be,Dyall:2006ec,Dyall:2016hn} with two additional diffuse functions for the halogens, and the Dunning aug-cc-pVTZ sets~\cite{Kendall:1992bj} for oxygen and hydrogen. Due to constraints in computational resources, for the [\ce{(F(H2O)8)-}@\ce{(H2O)_{42}}] partition DFT-in-DFT calculations were performed exclusively, using the DC Hamiltonian for \ce{F-}. In order to estimate the energies at the complete basis set (CBS) limit calculations with augmented quadruple-zeta basis sets were also performed: for \ce{F-} and \ce{Cl-} it was computationally feasible to do so for all snapshots. For the other halides this was not the case and estimates for the CBS energies were obtained based on quadruple-zeta calculation on the bare halides. The dataset comprising the DFT-in-DFT and CC-in-DFT calculation is available in the Zenodo repository~\cite{paper:dataset}. 

\begin{figure}[h]
    \centering
\begin{minipage}[b]{0.48\linewidth}
\centering
\includegraphics[width=\textwidth]{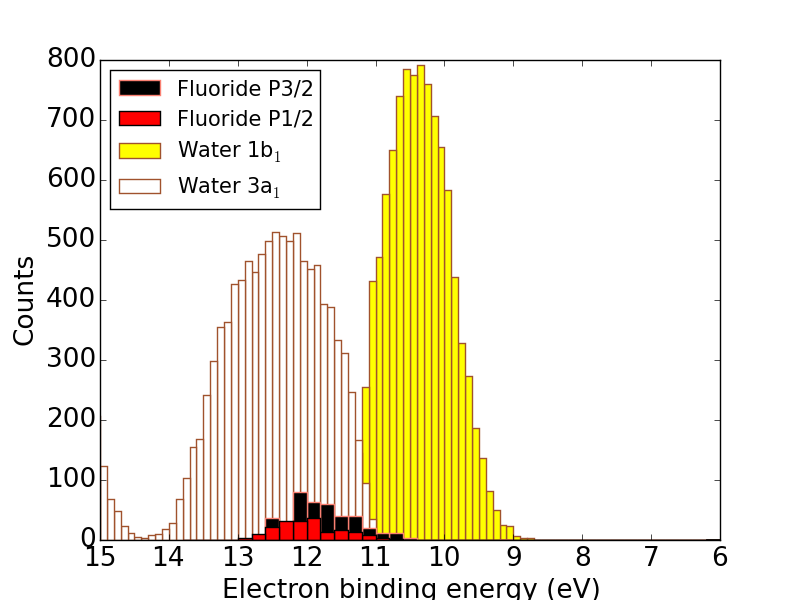}
\label{fig:fluoride}
{[\ce{F-}@\ce{(H2O)_{50}}]}
\end{minipage}
\begin{minipage}[b]{0.48\linewidth}
\centering
\includegraphics[width=\textwidth]{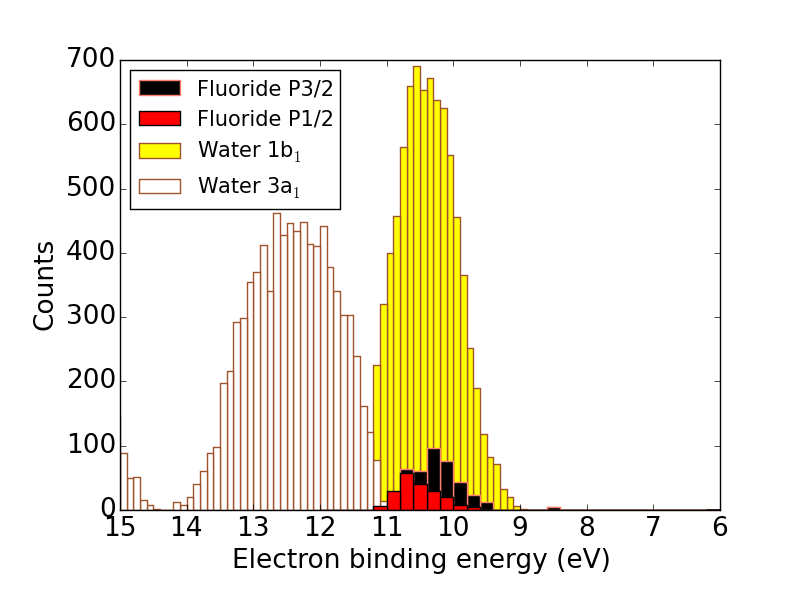}
\label{fig:fluoride8}
{[\ce{(F(H2O)8)-}@\ce{(H2O)_{42}}]}
\end{minipage}
\begin{minipage}[b]{0.48\linewidth}
\centering
\includegraphics[width=\textwidth]{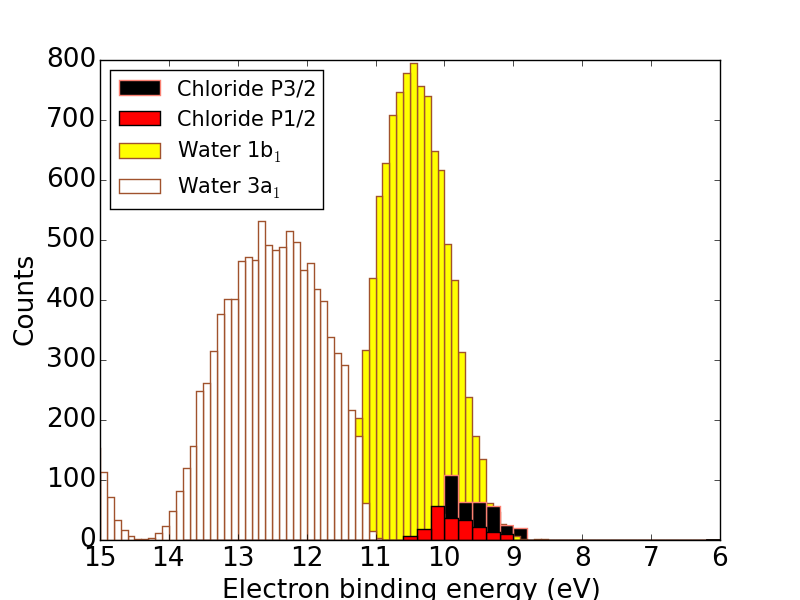}
\label{fig:chloride}
{[\ce{Cl-}@\ce{(H2O)_{50}}]}
\end{minipage}
\begin{minipage}[b]{0.48\linewidth}
\centering
\includegraphics[width=\textwidth]{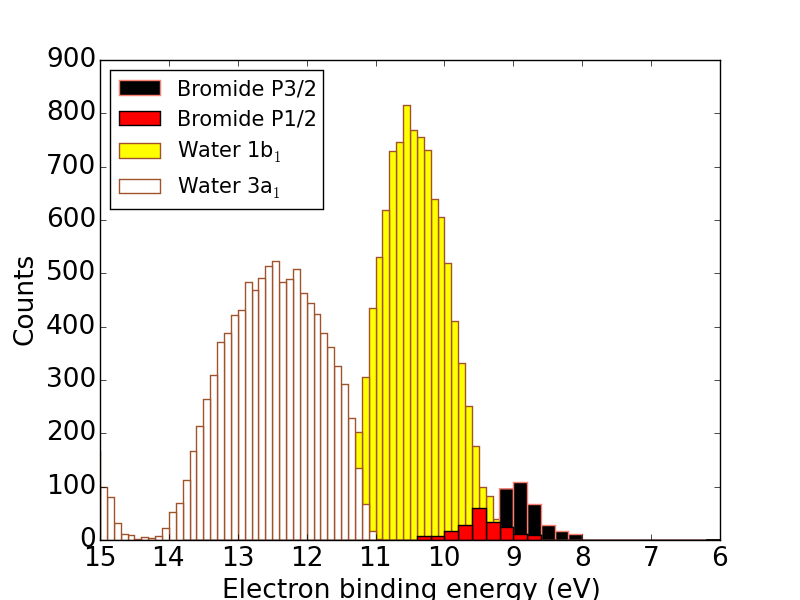}
\label{fig:bromide}
{[\ce{Br-}@\ce{(H2O)_{50}}]}
\end{minipage}
\begin{minipage}[b]{0.48\linewidth}
\centering
\includegraphics[width=\textwidth]{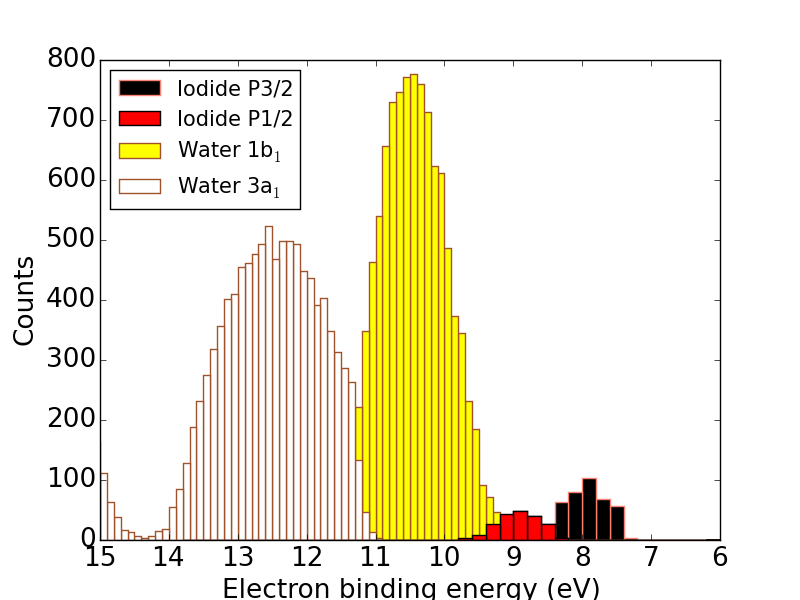}
\label{fig:iodide}
{[\ce{I-}@\ce{(H2O)_{50}}]}
\end{minipage}
\begin{minipage}[b]{0.48\linewidth}
\centering
\includegraphics[width=\textwidth]{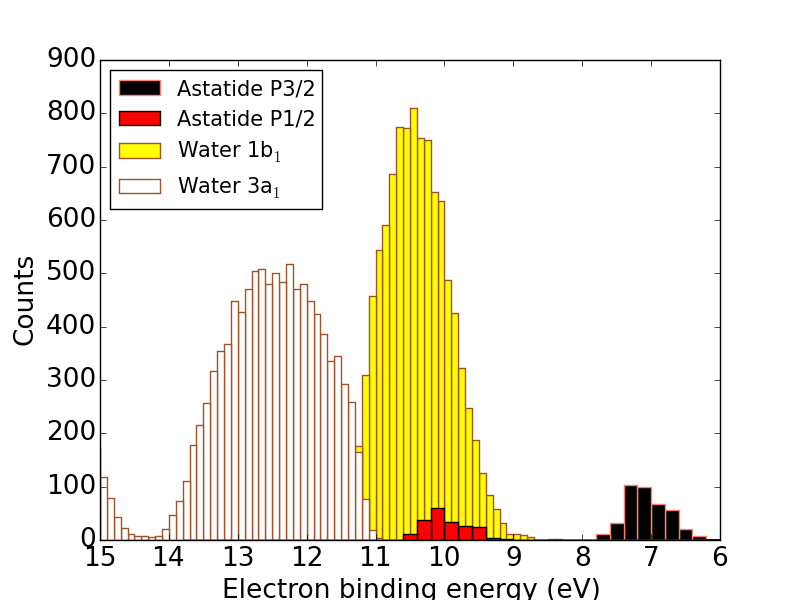}
\label{fig:astatide}
{[\ce{At-}@\ce{(H2O)_{50}}]}
\end{minipage}
    \caption{Electron binding energies spectra for the [\ce{X-}@\ce{(H2O)_{50}}] systems over the 100 snapshots. Halides BEs obtained with triple-zeta basis sets from DC EOM (except for [\ce{(F(H2O)8)-}@\ce{(H2O)_{42}}] obtained with DC SAOP)~\cite{paper:figures}.}
    \label{potential-energy-curves-xo-all}
\end{figure}

We start by discussing the trends along the series for the BEs over the 100 snapshots, presented in Fig.~\ref{potential-energy-curves-xo-all} as histograms plots, with the area under each rectangle being proportional to the number of BEs found at each energy interval. There is very little variation on the BEs of the water subsystems (the yellow and brown rectangles) upon changing the halogen.  For the halogens one finds, first, the displacement of the first ionization energy peak, which in the presence of SOC corresponds to the ${^2}P_{1/2}$ halogen atom ground electronic states, towards lower energies as the halogen gets heavier. This results in a clear separation between the halogen and water peaks from \ce{Br-} onwards. One can also see, as expected from experiments and prior calculations, that irrespective of the treatment of the first solvation shell of \ce{F-} (here only carried out with DC SAOP calculations as explained above), its electron BEs remain entangled with those of the water cluster. Second, the increasing separation between the ${^2}P_{1/2}$ and ${^2}P_{3/2}$ components of the halogen ground-state is clearly seen, and for \ce{I-} the two peaks are clearly distinguishable from those of the water. It is interesting to note, however, that for \ce{At-} the SOC effect is so large (with a ${^2}P_{3/2}$--${^2}P_{1/2}$ splitting of $\simeq$3.0 eV) that the ${^2}P_{1/2}$ peak ends up overlapping with those of water.

\begin{table}[h]
\caption{Average electron binding energies (BE, in eV) for the spin-orbit coupled components of the $P$ states of the hydrated halogens from EOM and SAOP (DC) calculations on the embedded halides with triple-zeta basis sets and the CBS values (*: estimates from single quadruple-zeta calculations); and water droplet valence bands from SAOP (SR-ZORA) calculations for the \ce{(H2O)_{50}} and \ce{(H2O)_{42}} subsystems.}
\begin{center}
\begin{tabular}{l  rr p{0.01cm} rr p{0.01cm} rr}
\hline
\hline
                  &     \multicolumn{5}{c}{Halogen}  &&    \multicolumn{2}{c}{Water}  \\
\cline{2-6}
\cline{8-9}
                             &      \multicolumn{2}{c}{BE$_{3/2}$}               && \multicolumn{2}{c}{BE$_{1/2}$}          &&  BE$_{1b_1}$   & BE$_{3a_1}$  \\
                             \cline{2-3}\cline{5-6}\cline{8-9}
 species                   &   EOM        &    SAOP     &&     EOM        &   SAOP         &&   \multicolumn{2}{c}{SAOP} \\
\hline
\multicolumn{9}{c}{triple-zeta bases}\\
\hline
\ce{F-}                    &  11.8(5)     &   11.4(5)   &&    12.0(5)     &   11.5(4)      &&   10.4(5)     &  12.4(7)     \\
\ce{F(H2O)8-}              &              &   10.3(4)   &&                &   10.5(3)      &&   10.4(5)     &  12.4(7)      \\
\ce{Cl-}                   &  9.7(3)      &   9.4(4)    &&    9.9(3)      &   9.5(4)       &&   10.4(5)     &  12.5(4)     \\
\ce{Br-}                   &  9.0(4)      &   8.7(3)    &&    9.5(4)      &   9.2(4)       &&   10.4(5)     &  12.5(4)     \\
\ce{I-}                    &  7.9(3)      &   7.8(3)    &&    8.9(3)      &   8.6(3)       &&   10.4(5)     &  12.5(4)     \\
\ce{At-}                   &  7.1(3)      &   7.0(3)    &&    10.0(3)     &   9.5(3)       &&   10.4(5)     &  12.5(4)     \\
\hline
\multicolumn{9}{c}{CBS (F$^-$, Cl$^-$) and CBS* (Br$^-$--At$^-$)}\\
\hline
\ce{F-}                    &  11.9(5)     &   11.4(5)   &&    12.1(5)     &   11.5(4)      &&               &             \\
\ce{F(H2O)8-}              &              &   10.3(4)   &&                &   10.5(3)      &&               &             \\
\ce{Cl-}                   &  9.9(3)      &   9.4(4)    &&    10.1(3)     &   9.5(4)       &&               &             \\
\ce{Br-}                   &  9.0(4)      &   8.7(3)    &&    9.5(4)      &   9.2(4)       &&               &             \\
\ce{I-}                    &  8.0(3)      &   7.8(3)    &&    9.0(3)      &   8.6(3)       &&               &             \\
\ce{At-}                   &  7.1(3)      &   7.0(3)    &&    10.1(3)     &   9.5(3)       &&               &             \\
\hline
\hline
\end{tabular}
\end{center}
\label{tab:FDE-IPs}
\end{table}

\begin{table}[h]
\caption{Experimental electron binding energies (BE, in eV)  for the spin-orbit coupled components of the $P$ states of the solvated halide and bulk water valence bands from (a) Kurahashi and coworkers~\cite{Kurahashi:2014df}; (b) Winter and coworkers~\cite{Winter:2005kx} ($^\dagger$ average value of the 3$a_1$ H and 3$a_1$ L bands; $^\ddag \Omega=3/2$; $^\ast \Omega=1/2 $).}
\begin{center}
\begin{tabular}{l  rr p{0.01cm} rr p{0.01cm} rr}
\hline
\hline
                  &     \multicolumn{2}{c}{Halogen}  &&    \multicolumn{5}{c}{Water} \\
\cline{2-3}
\cline{5-9}
                             &	     \multicolumn{2}{c}{BE$_p$}	          &&  \multicolumn{2}{c}{BE$_{1b_1}$}   &&  \multicolumn{2}{c}{BE$_{3a_1}$}  \\
                            \cline{2-3}\cline{5-6}\cline{8-9}
 species    &     (a)   &   (b) 
                 &&  (a)    &   (b)   
                 &&  (a)    &   (b)      \\ 
\hline
\ce{F-}        &	9.8	        &                        &&                 &            &&            &     \\
\ce{Cl-}	&	9.5(2)       & 	9.60(7)      &&                 &            &&            &     \\
\ce{Br-}      &	9.00(7)	&	8.80(6)      &&                 &            &&            &      \\
                &	                    &        8.1(1)      &&                 &            &&            &      \\
\ce{I-}         &     8.03(6)$^\ddag$  &   7.7(2)$^\ddag$ &&   11.31(4)      &  11.16(4)  &&  13.78(7)$^\dagger$   & 13.50(10)   \\
                &     8.96(7)$^\ast$	 &	8.8(2)$^\ast$    &&                      &                 &&                                     &   \\
\hline
\hline
\end{tabular}
\end{center}
\label{tab:experimental-IPs}
\end{table}

\begin{table}
\caption{Gas-phase electron binding energies (BE, in eV) for the halides (DC) and the water molecule (SR-ZORA, PBE optimized geometry) ($\dagger$ : CCSD(T)).}
\begin{tabular}{llrr p{0.01cm} rrl}
\hline\hline
Species  &   & \multicolumn{2}{c}{SAOP}    && \multicolumn{2}{c}{EOM} &  \\
\cline{3-4}
\cline{6-7}
         &             & triple-zeta & CBS && triple-zeta &  CBS & Exp.\ \\
\hline
\ce{F-}  & BE$_{3/2}$  &    3.16 &    3.16 &&    3.32 &     3.45 &   3.40~\cite{IP-Popp-ZN1967-A-22-254,IP-Milstein-JCP1971-55-4146}  \\
\ce{Cl-} & BE$_{3/2}$  &    3.41 &    3.41 &&    3.59 &     3.77 &   3.62~\cite{IP-Muck-ZN1968-23-1213,IP-McDermid-JPC1983--C7-461}  \\
\ce{Br-} & BE$_{3/2}$  &    3.23 &    3.23 &&    3.40 &     3.48 &   3.37~\cite{IP-Berry-JCP1963-38-1540,IP-Frank-ZN1970-25-1617}  \\
\ce{I-}  & BE$_{3/2}$  &    3.02 &    3.02 &&    3.12 &     3.19 &   3.06~\cite{IP-Webster-JCP1983-78-646}  \\
\ce{At-} & BE$_{3/2}$  &    2.48 &    2.48 &&    2.41 &     2.55 &   2.40$^\dagger$~\cite{IP-Borschevsky-PR2015-91-020501} \\
H$_2$O   & BE$_{1b_1}$ &   12.33 &         &&         &          &  12.62~\cite{water-Reutt-JCP1986-85-6928} \\
\hline\hline
\end{tabular}
\label{Tab:IP-atomic-CBS}
\end{table}

Table~\ref{tab:FDE-IPs} summarizes the average BEs for the DFT-in-DFT and CC-in-DFT calculations of Fig.~\ref{potential-energy-curves-xo-all} (corresponding to peak maxima), while the experimental results are shown in Table~\ref{tab:experimental-IPs}. By their comparison one sees that, apart from the \ce{F-} case, the EOM results agree rather well with the experimental peak maxima for the halides, with differences of about \SI{0.2}{\eV} for \ce{Cl-} and about \SI{0.1}{eV} for \ce{Br-} and \ce{I-}. 
We attribute this relative improvement along the series to a decrease in entanglement between halide and the surrounding water molecules as the halide gets heavier~\cite{Real:2016bhb}, which would make our simple embedding model better represent the physical system. For \ce{I-}, the only system for which Kurahashi and coworkers~\cite{Kurahashi:2014df} provide the spin-orbit splitting of the ${^2}P$ state, there is also very good agreement with experiment for the ionization from the ${^2}P_{1/2}$ state. 

Table~\ref{tab:FDE-IPs} presents results for the halides obtained with triple-zeta bases and CBS energies (for \ce{F-} and \ce{Cl-}) estimates (for \ce{Br-} to \ce{At-}). A comparison of EOM triple-zeta and CBS results indicates the latter show a discrete improvement over the former, and in general make our results closer to experiment. Furthermore, the SAOP results are in rather good agreement with the EOM values, with rather systematic differences in the order of \SI{0.4}{eV}. This underscored the good performance of SAOP for BEs, especially in view of its modest computational cost, and validates our choice of employing SAOP for the DFT-in-DFT calculations. Additionally, as seen from Table~\ref{Tab:IP-atomic-CBS}, SAOP and EOM yield good gas-phase BEs, meaning the experimental halide BE shifts upon solvation is well-reproduced. 
That said, our embedding model shows what appears to be a systematic underestimation of the water spectra, by roughly \SI{1}{eV} for the $b_1$ and $a_1$ peaks. Part of this discrepancy 
should originate from using SAOP rather than EOM energies (if errors follow those for the halides discussed above, up to 0.4--0.5 eV).
We believe the other major source of errors is the discrete size of the droplets used, since the experimental results are for bulk water, and intend to investigate this issue in a subsequent publication. 

\begin{table}[h]
\caption{Selected theoretical electron binding energies (BE, in eV) from the literatures for solvated \ce{F-} and \ce{Cl-} using the
$G_0W_0$~\cite{Gaiduk:2016cf} approach, and the Outer-Valence Green’s Function (OGVF), Partial third order (P3) and renormalized Partial third order (P3+) propagator approaches combined with PCM (polarizable continuum model)~\cite{Dolgounitcheva:2012ik} or explicit solvation (PC: point-charge embedding)~\cite{fluoride-Canuto-JCP2010-132-214507}.} 
\begin{center}
\begin{tabular}{l  rr p{0.01cm} rr p{0.01cm} rr}
\hline
\hline
Method		&	Cl$^-$	&	F$^-$	\\
\hline			
$G_0W_0$/PBE~\cite{Gaiduk:2016cf}		&	8.76	&		\\
$G_0W_0$/PBE0~\cite{Gaiduk:2016cf}		&	9.43	&		\\
$G_0W_0$/RSH~\cite{Gaiduk:2016cf}		&	9.86	&		\\
$G_0W_0$/sc-hybrid~\cite{Gaiduk:2016cf}	&	9.89	&		\\

OGVF/PCM~\cite{Dolgounitcheva:2012ik}		&	10.53	&	10.70	\\
P3/PCM~\cite{Dolgounitcheva:2012ik}		&	10.32	&	12.21	\\
P3+/PCM~\cite{Dolgounitcheva:2012ik}		&	10.29	&	12.02	\\

P3/6H$_2$O~\cite{fluoride-Canuto-JCP2010-132-214507}               &    6.95       &  \\
P3/6H$_2$O + 60H$_2$O(PC)~\cite{fluoride-Canuto-JCP2010-132-214507} &    9.41       &  \\
\hline					
\hline						
\end{tabular}
\end{center}
\label{tab:other-theoretical-IPs}
\end{table}

For \ce{Cl-} a comparison to prior theoretical results can be made to the $G_0W_0$ calculations (without SOC) of Gaiduk and coworkers~\cite{Gaiduk:2016cf}, shown in Table~\ref{tab:other-theoretical-IPs}, for which the most sophisticated calculation using the sc-hybrid density functional places the peak position at \SI{9.89}{\eV}. This is higher than the experimental results by a little over \SI{0.3}{\eV}. It is also higher than EOM calculations, even if it is compared to our ${^2}P$ term value of \SI{9.76}{\eV}. The $G_0W_0$/sc-hybrid calculations show very good agreement with experiment for the water peaks, though a comparison to our results would be somewhat biased since the $G_0W_0$ ones are made for a bulk liquid and ours not. It is important to note the $G_0W_0$ results do not show a very good agreement with the experimental BEs if less sophisticated functionals such as PBE and PBE0 are used--in fact, the DC SAOP results are of slightly better quality than those.

Another relevant comparison is with electron propagator calculations of Dolgounitcheva and coworkers~\cite{Dolgounitcheva:2012ik}, performed for microsolvated clusters of \ce{F-} and \ce{Cl-}, and included the effect of outer solvation shells via PCM. For \ce{Cl-} the propagator results agree well with each other but are nevertheless 0.7 to \SI{1}{\eV} higher than experiment, whereas our results are not more than \SI{0.2}{\eV} higher. For the first ionization of \ce{F-} to which there are significant contributions from Dyson orbitals on \ce{F}, the propagator results are closer to each other but again quite far from experiment.  If part of the discrepancy comes from differences in treatment of electron correlation between the propagators and EOM (or SAOP) and basis set effects (bases smaller than ours were used), the most significant contribution shoud be due to the explicit inclusion of the outer solvation shells in our calculations. The importance of this effect is seen in the P3 calculations of Canuto and coworkers~\cite{fluoride-Canuto-JCP2010-132-214507} which, when considering outer-shell effects via point-charge embedding, recover nearly \SI{2.5}{\eV} with respect to the microsolvated ion, showing an agreement to experiment similar to SAOP.

In conclusion, our results show FDE is a viable method for obtaining quantitatively accurate electron binding energies (and with that simulate PE spectra) in the valence region for species in solution. For systems not undergoing chemical changes, the combination of CC-in-DFT calculations with CMD simulations with polarizable force fields can yield results which rival much more sophisticated simulation approaches but at a much smaller computational cost (the embedded EOM calculations take about a day per snapshot on 4 cores for \ce{At-}, the most expensive calculations). In this sense, the SAOP model potential appears as a rather interesting alternative to more computationally expensive functionals by itself or, eventually, being combined with many-body treatments based on the $GW$ method. Finally, our work was based on droplet simulations, which can be interesting to investigate systems made up by a relatively small amount of water molecules, though monitoring droplet size effects on such properties and their convergence towards the bulk requires further investigations. The FDE calculations are, however, completely agnostic to the nature of the procedure employed to obtain the structures, and can be equally applied to snapshots from standard (or FDE-based~\cite{Genova:2017dn}) CPMD calculations (whenever DFT-based interaction potentials are sufficiently accurate~\cite{water-Gillan-JCP2016-144-130901}) or static band-structure FDE calculations~\cite{Toelle:2018qua} that naturally describe long-range interactions in extended systems.  

We acknowledge support from the Labex CaPPA (Chemical and Physical Properties of the Atmosphere, contract ``ANR-11-LABX-0005-01''),
CPER CLIMIBIO (European Regional Development Fund, Hauts de France council, French  Ministry of Higher Education and Research), CNRS Institute of Physics (PICS grant 6386)
and French national supercomputing facilities (grant DARI A0030801859).

\end{document}


\title{Supplementary Information \\ Predictive simulations of ionization energies of solvated halide ions with relativistic embedded Equation of Motion Coupled-Cluster Theory}

\author{Yassine Bouchafra}
\affiliation{Universit\'e de Lille, CNRS, UMR 8523 -- PhLAM -- Physique des Lasers, Atomes et Mol\'ecules, F-59000 Lille, France Tel: +33-3-2043-4163}

\author{Avijit Shee}
\affiliation{Department of Chemistry, University of Michigan, 930 N.\ University, Ann Arbor, MI 48109-1055, USA}
\altaffiliation{Universit\'e de Lille, CNRS, UMR 8523 -- PhLAM -- Physique des Lasers, Atomes et Mol\'ecules, F-59000 Lille, France Tel: +33-3-2043-4163}

\author {Florent R\'eal}
\affiliation{Universit\'e de Lille, CNRS, UMR 8523 -- PhLAM -- Physique des Lasers, Atomes et Mol\'ecules, F-59000 Lille, France; Tel: +33-3-2043-4163}
\author {Val\'erie Vallet}
\affiliation{Universit\'e de Lille, CNRS, UMR 8523 -- PhLAM -- Physique des Lasers, Atomes et Mol\'ecules, F-59000 Lille, France; Tel: +33-3-2043-4163}
\author{Andr\'e Severo Pereira Gomes}
\email{andre.gomes@univ-lille.fr (corresponding author)}
\affiliation{Universit\'e de Lille, CNRS, UMR 8523 -- PhLAM -- Physique des Lasers, Atomes et Mol\'ecules, F-59000 Lille, France; Tel: +33-3-2043-4163}

\date{\today}
\revised{\today}

\maketitle


\section{Atomic calculations}

\begin{table}
\caption{Gas-phase atomic binding energies (BE, in eV) computed at the SAOP and EOM (DC) levels for triple-zeta, quadruple-zeta basis sets and at the CBS levels.}
\begin{tabular}{ll*7S}
\hline
\hline
Species & & \multicolumn{3}{c}{SAOP} && \multicolumn{3}{c}{EOM}\\
\cline{3-5}\cline{7-9}
         &            & TA   & QZ & CBS && TZ & QZ & CBS\\
\hline
\ce{F-}  & BE$_{3/2}$ & 3.16 & 3.16 & 3.16 &&  3.32 & 3.39 & 3.45 \\
         & BE$_{1/2}$ & 3.21 & 3.21 & 3.21 &&  3.37 & 3.45 & 3.51 \\
\ce{Cl-} & BE$_{3/2}$ & 3.41 & 3.41 & 3.41 &&  3.59 & 3.69 & 3.77 \\
         & BE$_{1/2}$ & 3.51 & 3.51 & 3.50 &&  3.70 & 3.81 & 3.89 \\
\ce{Br-} & BE$_{3/2}$ & 3.23 & 3.23 & 3.23 &&  3.40 & 3.45 & 3.48 \\
         & BE$_{1/2}$ & 3.65 & 3.65 & 3.65 &&  3.89 & 3.93 & 3.96 \\
\ce{I-}  & BE$_{3/2}$ & 3.02 & 3.02 & 3.02 &&  3.12 & 3.16 & 3.19 \\
         & BE$_{1/2}$ & 3.87 & 3.87 & 3.87 &&  4.09 & 4.14 & 4.18 \\
\ce{At-} & BE$_{3/2}$ & 2.48 & 2.48 & 2.48 &&  2.41 & 2.49 & 2.55 \\
         & BE$_{1/2}$ & 5.05 & 5.05 & 5.05 &&  5.35 & 5.44 & 5.51 \\
\hline
\hline
\end{tabular}
\label{Tab:IP-atomic-CBS}
\end{table}

Complete Basis Set (CBS) values are calculated using the following formula, in which $n$ is the basis set cardinal number:
\begin{equation}
E=E_{cbs}+\frac{A}{n^3} 
\end{equation}
Thus for two cardinal numbers $n_1=3$ (triple-zeta) and $n_2=4$  (quadruple-zeta), one can write
\begin{align}
E(n_1) &= E_{cbs}+\frac{A}{n_1^3} \\
E(n_2) &= E_{cbs}+\frac{A}{n_2^3} 
\end{align}
leading to the CBS extrapolated energy $E_{cbs}$:
\begin{equation}
E_{cbs}=\frac{E(n_1)n_1^3-E(n_2)n_2^3}{n_1^3-n_2^3} 
\end{equation}

\section{Choice of the embedding model for the DFT-in-DFT FnT calculations}
\subsection{Molecular orbitals compositions in the  \ce{[X(H2O)_{50}]-} calculations}

\begin{figure}[ht]
  \centering
  \includegraphics[width=0.90\textwidth]{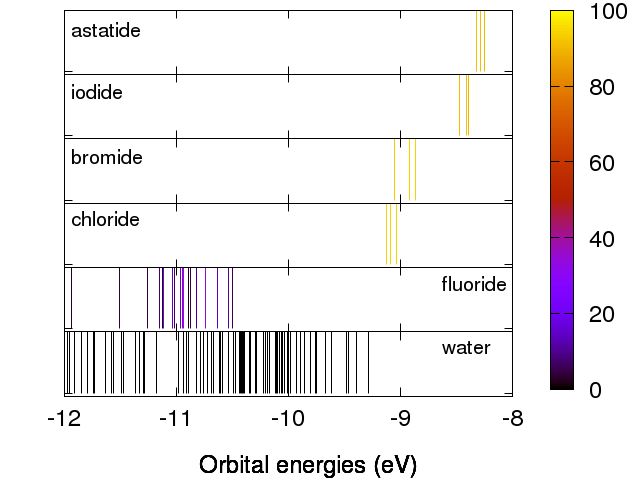}
\caption{Contributions of the halide valence p orbitals into the molecular orbitals of \ce{[X(H2O)_{50}]-} hydrated cluster for a single snapshot, from 0 (black) up to 40\% for fluoride, and 100\% for the heavier halides. All calculations refer to scalar relativistic ZORA SAOP calculations.~\cite{paper:figures}}
\label{Fig:X-MOcontribution}
\end{figure}

From the scalar-relativistic ZORA SAOP calculations on the \ce{[X(H2O)_{50}]-} supermolecular systems we have drawn in Figure~\ref{Fig:X-MOcontribution} the percentage contribution of the halide valence p orbitals into each molecular orbital. For all halides heavier than fluoride, the three highest occupied molecular orbitals correspond to the valence p halide orbitals, while for fluoride its 2p orbitals are immersed into the water valence manifold.

\subsection{Influence of the embedding model on binding energies}

The subsystem DFT  approach~\cite{Cortona:1991fm,Cortona:1992br,Wesolowski:1993cv} invokes calculation of the effective embedding potential, in order to take into account the effect of the environment on the embedded system. The simplest implementation of subsystem DFT is frozen density embedding (FDE)~\cite{Wesolowski:1993cv}, in which the environment subsystem density $n_\text{II}(r)$ is kept frozen while the total energy is minimized with respect to changes in the other subsystem density $n_\text{I}(r)$. The minimization of the total energy  with respect to the supermolecular density can be achieved through freeze-and-thaw (FnT) cycles (typically less than 20), where the roles of the subsystems I and II are iteratively interchanged. The relaxation steps are needed to account for the deformation/polarization of the subsystem's densities, in the presence of charges. 

\begin{figure}[ht]
    \centering
     \begin{subfigure}[b]{0.32\textwidth} 
       \centering \includegraphics[width=\textwidth]{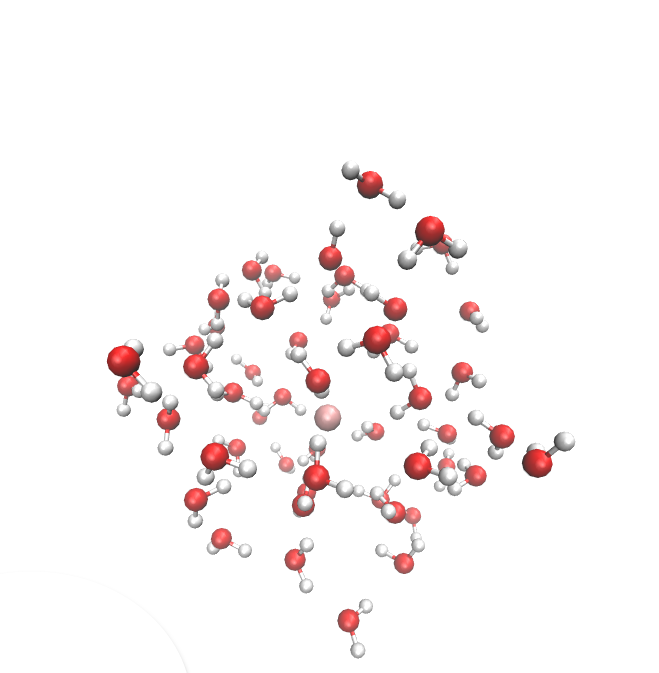}
       \caption{\ce{[I(H2O)_{50}]-}}\label{Fig:supermolecule}
    \end{subfigure}
    \begin{subfigure}[b]{0.32\textwidth}
        \centering \includegraphics[width=\textwidth]{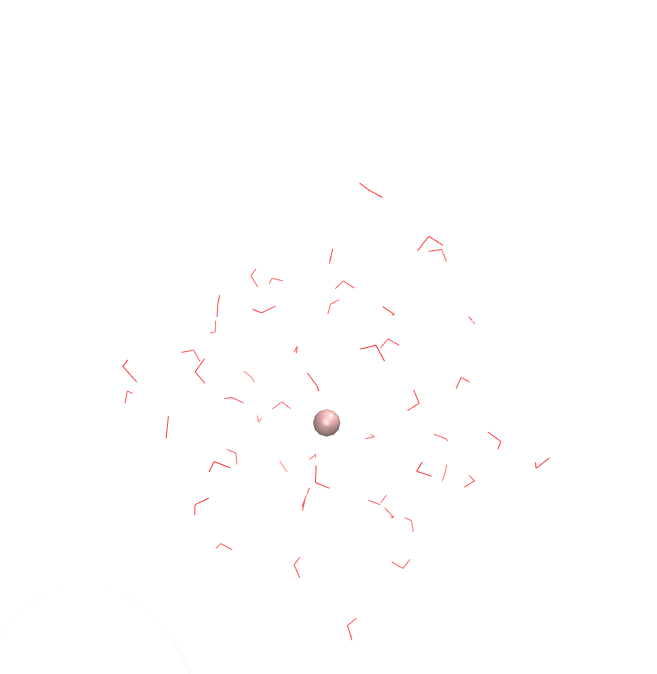}
        \caption{[\ce{I-}@\ce{(H2O)_{50}}]}\label{Fig:I_0_50}
    \end{subfigure}
    \begin{subfigure}[b]{0.32\textwidth}
        \centering \includegraphics[width=\textwidth]{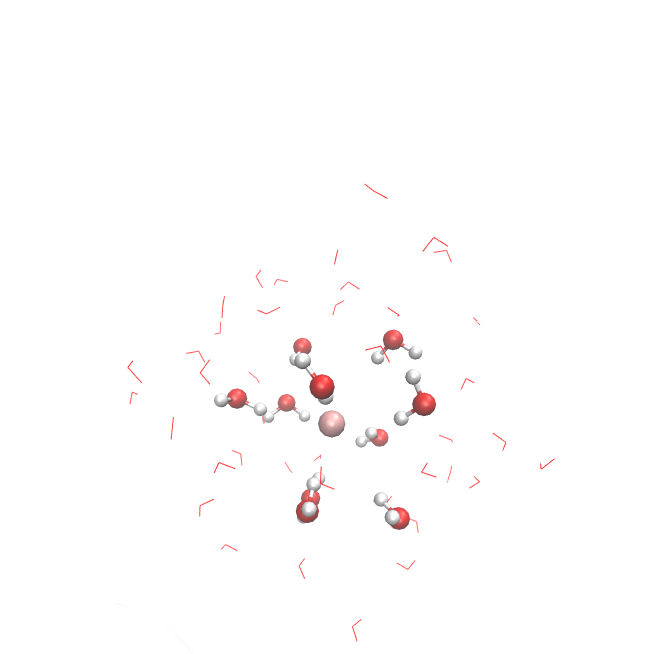}
        \caption{[\ce{I(H2O)_{10}-}@\ce{(H2O)_{40}}]}\label{Fig:I_10_40}
    \end{subfigure}
    \caption{Perspective views of the three different models, (a) the supermolecule, (b) iodide embedded in 50 water molecules; (c) \ce{I(H2O)_{10}-} embedded in 40 water molecules.~\cite{paper:figures}}\label{Fig:Subsystems}
\end{figure}

For this study, we have explored several density partitioning for the case of iodide hydrated by 50~water molecules, referred to as the supermolecule (cf. Figure~\ref{Fig:supermolecule}). The simplest embedded model includes the halide anion as the central subsystem, and the solvating water molecules in the environment as depicted in Figure~\ref{Fig:I_0_50}. A midway model places 10 water molecules constituting the iodide's first hydration shell, the 40 other water molecules being part of the embedding, as illustrated by Figure~\ref{Fig:I_10_40}.

\begin{figure}[ht]
\centering
\includegraphics[width=0.9\textwidth]{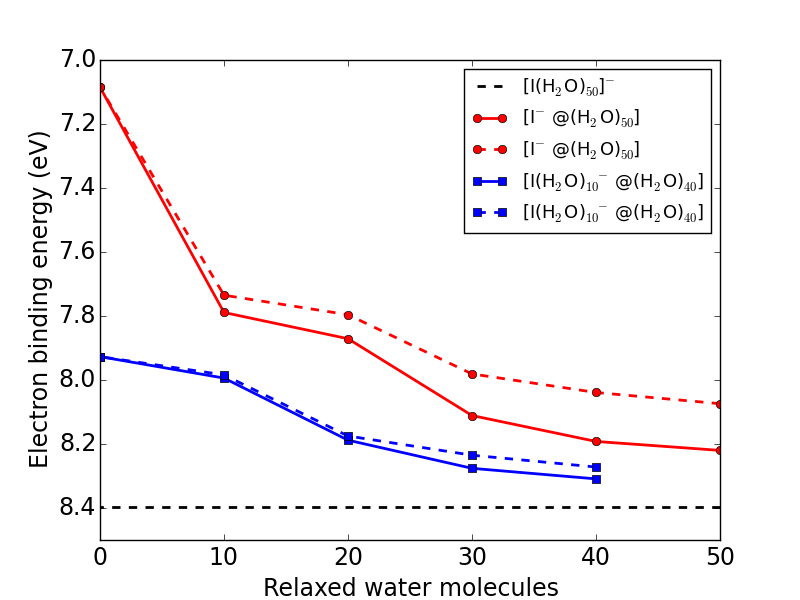}
\caption{Variation of the electron binding energy (SR-ZORA SAOP HOMO energy) for iodide embedded in water ([\ce{I-}@\ce{(H2O)_{50}}] (in red) and [\ce{I(H2O)_{10}-}@\ce{(H2O)_{40}}]) (in blue) as a function of the number of water molecules which have their density relaxed in the DFT-in-DFT procedure. The dashed lines correspond to FnT calculations in which the water molecules are considered individually, while for the plain lines correspond to the calculations in which the whole \ce{(H2O)_n} water cluster's density is relaxed. The horizontal black dashed line represents the supermolecule calculation.~\cite{paper:figures}}\label{fig:HOMO_I-_SM_vs_2_FDE}
\label{Fig:relaxation-test}
\end{figure}

Figure~\ref{Fig:relaxation-test} monitors the electron binding energy (SR-ZORA-SAOP) as a function of the density relaxation of the water molecules as a function of their distance from the iodide. For the [\ce{I-}@\ce{(H2O)_{50}}] model , the binding energy (BE) approaches the supermolecule value, when the 50 water molecules are relaxed as individual fragments (red dashed curve). The difference to the supermolecule reference is narrowed down to \SI{0.2}{\eV} when the 50 water molecules are relaxed altogether as a single fragment, certainly because the inaccuracies in the embedding potential due to the approximate kinetic energy functionals are minimized as the number of interacting fragment is kept small. It is noteworthy that including explicitly the first hydration shell of 10 water molecules in the central subsystem (blue curves), helps gaining only \SI{0.1}{\eV} with respect to the supermolecule at the expense of larger computational costs for the DFT-in-DFT FnT steps, and breaking the current computational limits of DC EOM-IP-CC calculations. This, together with the molecular orbital analysis supports the choice for all halogen heavier than fluoride of an embedding model composed of the halogen, embedded in a cluster of 50 water molecules. 

Fluoride stands out, as its valence 2p orbitals strongly overlap with water $1b_1$ and $3a_2$ water orbitals. In the [\ce{F-}@\ce{(H2O)_{50}}] model, the FnT relaxation steps cannot account for this strong entanglement, leading to overshoot the average BE of fluoride (\SI{11.4\pm0.5}{\eV}) by about \SI{1.6}{\eV}, with respect to the experimental value~\cite{Kurahashi:2014df}. It thus deemed necessary to treat fluoride with its first hydration shell built from the 8 closed water molecule, embedded in the other 42 molecules, to reach good agreement (\SI{10.3\pm0.4}{eV})  with experiment.

\begin{figure}[ht]
\centering

   \begin{subfigure}[b]{0.48\textwidth} 
       \centering \includegraphics[width=\textwidth]{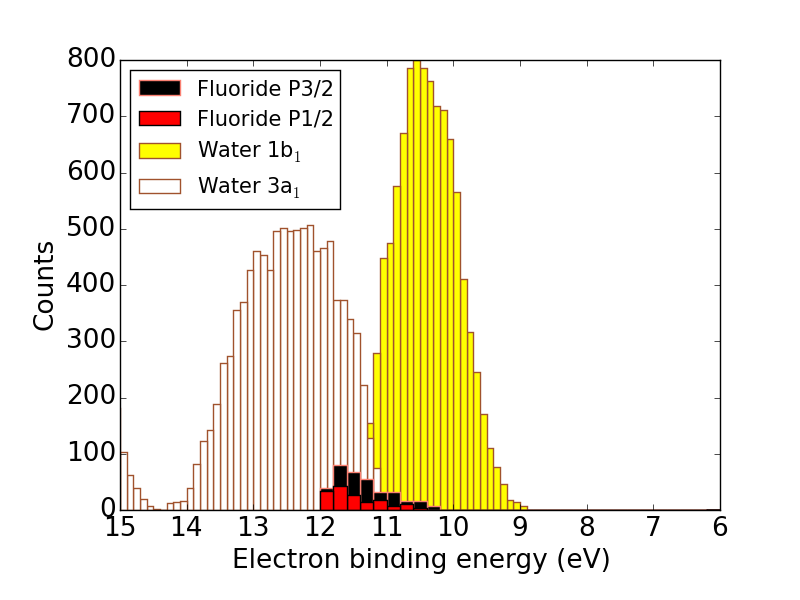}
       \caption{[\ce{F-}@\ce{(H2O)_{50}}]}\label{Fig:F-SAOP_0_50}
   \end{subfigure}
   \begin{subfigure}[b]{0.48\textwidth} 
       \centering \includegraphics[width=\textwidth]{Valence_photoemission_spectrum_F-_SO_DFT-in-DFT_6-15eV_8_42_100g.png}
       \caption{[\ce{(F(H2O)8)-}@\ce{(H2O)_{42}}]}\label{Fig:F-SAOP_8_42}
   \end{subfigure}

\caption{Comparison of the fluoride binding energies in \si{\eV} for the two models, [\ce{F-}@\ce{(H2O)_{50}}] and [\ce{(F(H2O)8)-}@\ce{(H2O)_{42}}], computed at the DC SAOP level for  100 snapshots.~\cite{paper:figures}}
\label{Fig:F-SAOP-model}
\end{figure}

\section{Time averages and correlation times}
\begin{figure}[ht]
\centering
   \begin{subfigure}[b]{0.48\textwidth} 
       \centering \includegraphics[width=\textwidth]{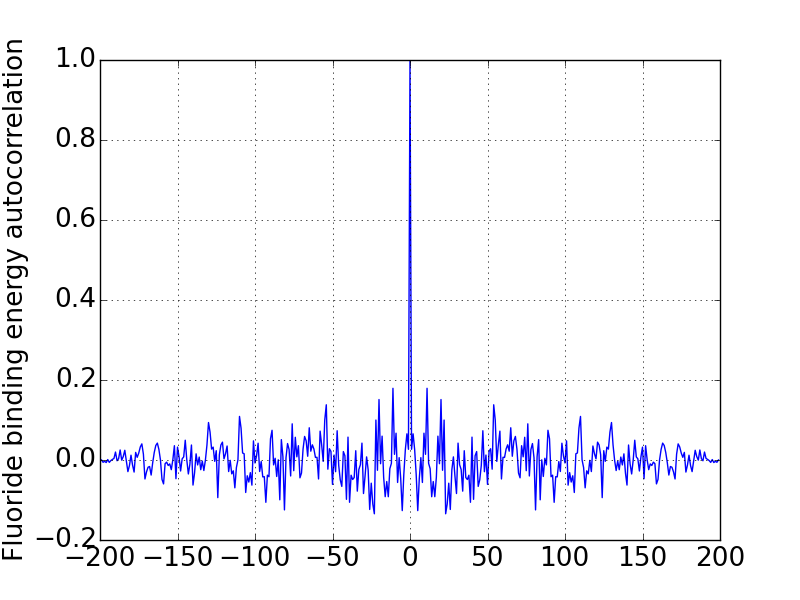}
       \caption{\ce{F-}}\label{Fig:F-BEautocorr}
   \end{subfigure}
   \begin{subfigure}[b]{0.48\textwidth} 
       \centering \includegraphics[width=\textwidth]{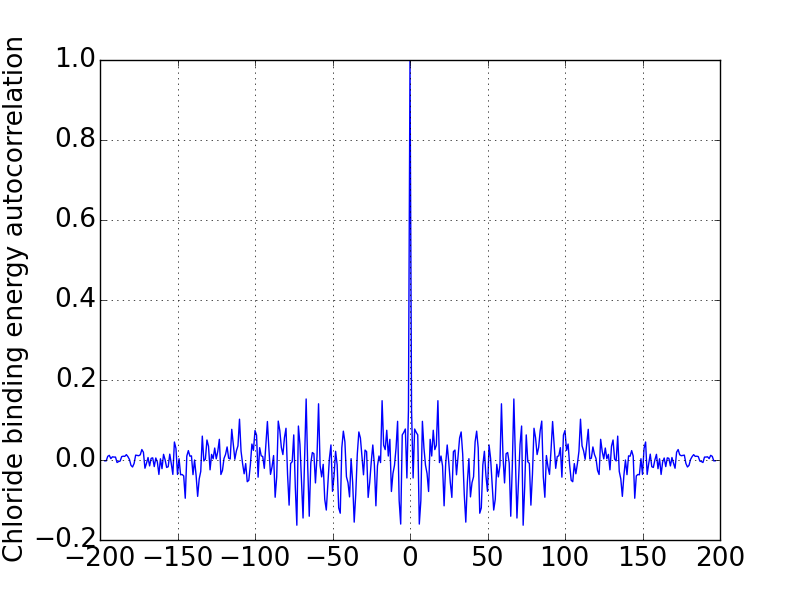}
       \caption{\ce{Cl-}}\label{Fig:Cl-BEautocorr}
   \end{subfigure}

   \begin{subfigure}[b]{0.48\textwidth} 
       \centering \includegraphics[width=\textwidth]{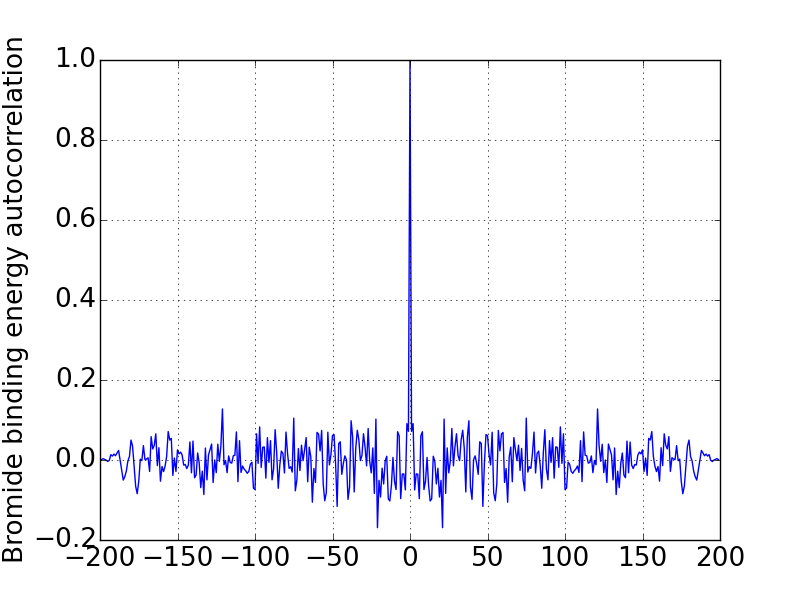}
       \caption{\ce{Br-}}\label{Fig:Br-BEautocorr}
   \end{subfigure}
   \begin{subfigure}[b]{0.48\textwidth} 
       \centering \includegraphics[width=\textwidth]{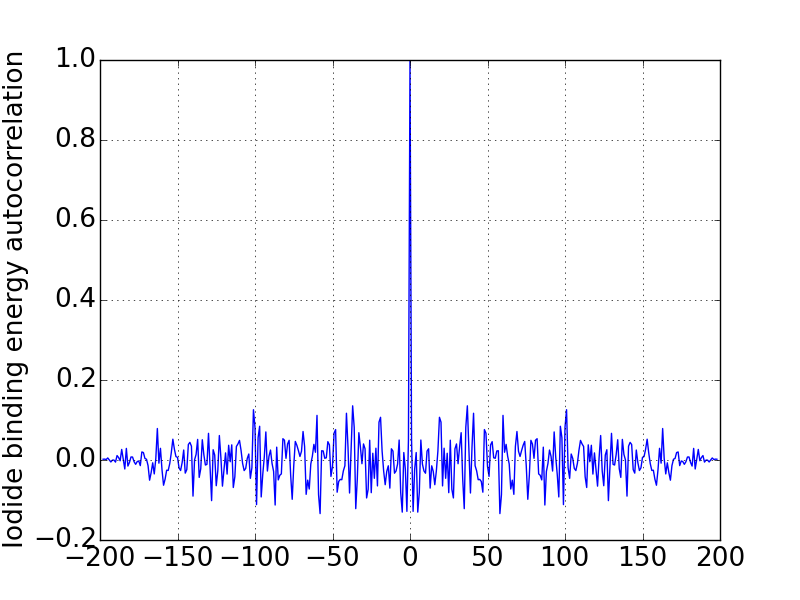}
       \caption{\ce{I-}}\label{Fig:I-BEautocorr}
   \end{subfigure}

   \begin{subfigure}[b]{0.48\textwidth} 
       \centering \includegraphics[width=\textwidth]{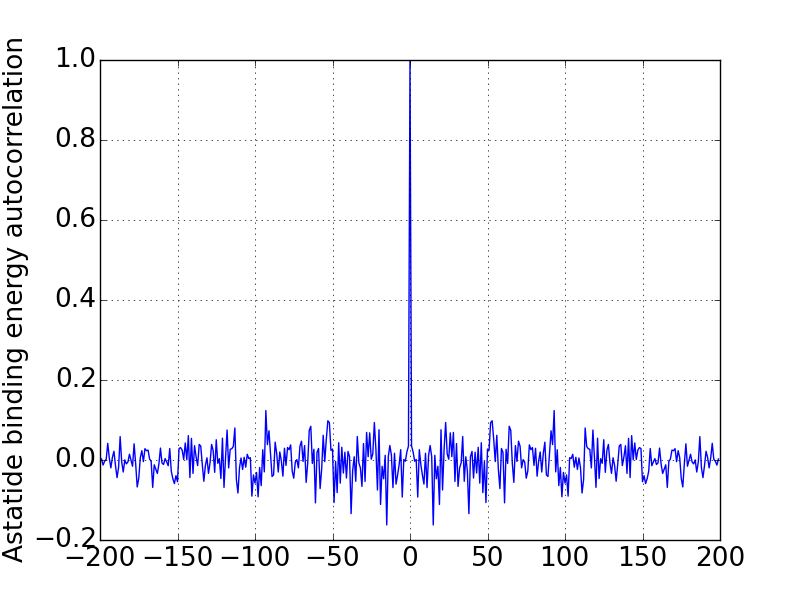}
       \caption{\ce{At-}}\label{Fig:At-BEautocorr}
   \end{subfigure}

\caption{Autocorrelation functions of the first halide binding energies computed at the DC SAOP level for the [\ce{X-}@\ce{(H2O)_{50}}] FnT models, computed for the 200 selected snapshots.~\cite{paper:figures}}
\label{Fig:X-BEautocorr}
\end{figure}

\begin{figure}[ht]
  \centering
   \begin{subfigure}[b]{0.48\textwidth} 
       \centering \includegraphics[width=\textwidth]{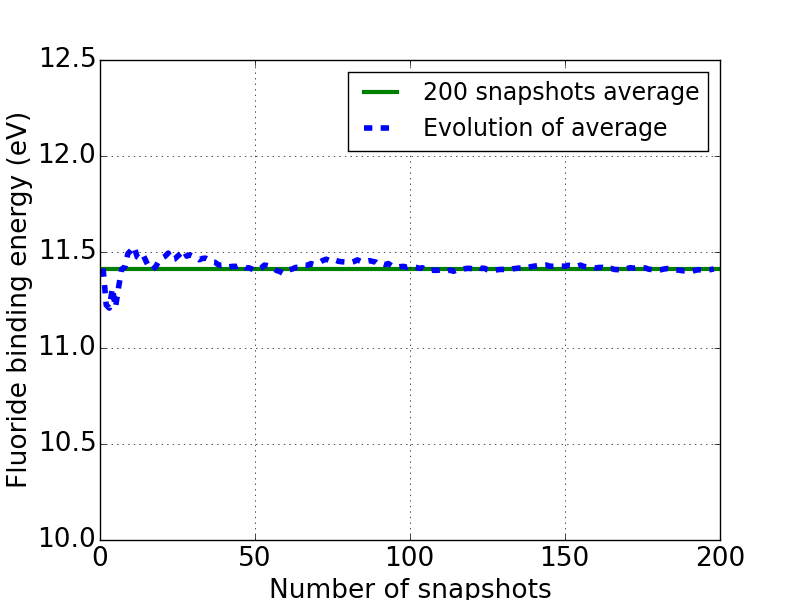}
       \caption{\ce{F-}}\label{Fig:F-BEevolution}
   \end{subfigure}
   \begin{subfigure}[b]{0.48\textwidth} 
       \centering \includegraphics[width=\textwidth]{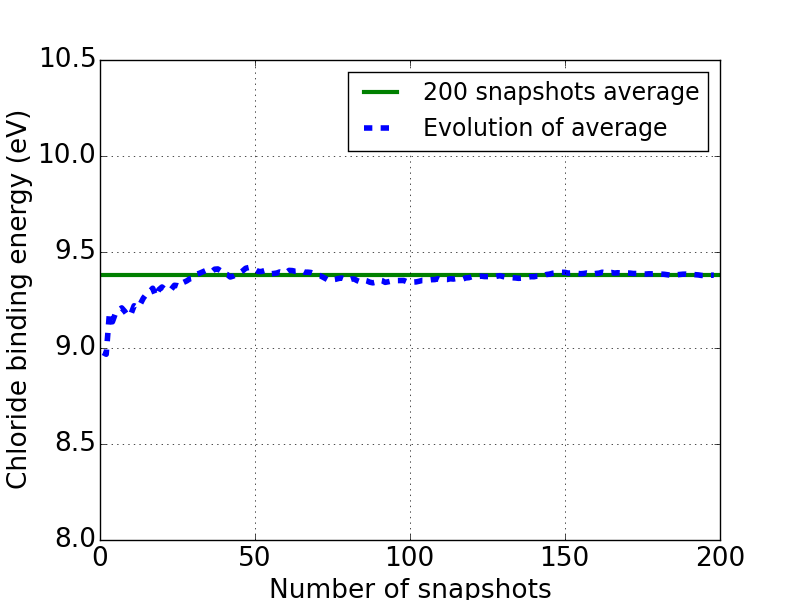}
       \caption{\ce{Cl-}}\label{Fig:Cl-BEevolution}
   \end{subfigure}

   \begin{subfigure}[b]{0.48\textwidth} 
       \centering \includegraphics[width=\textwidth]{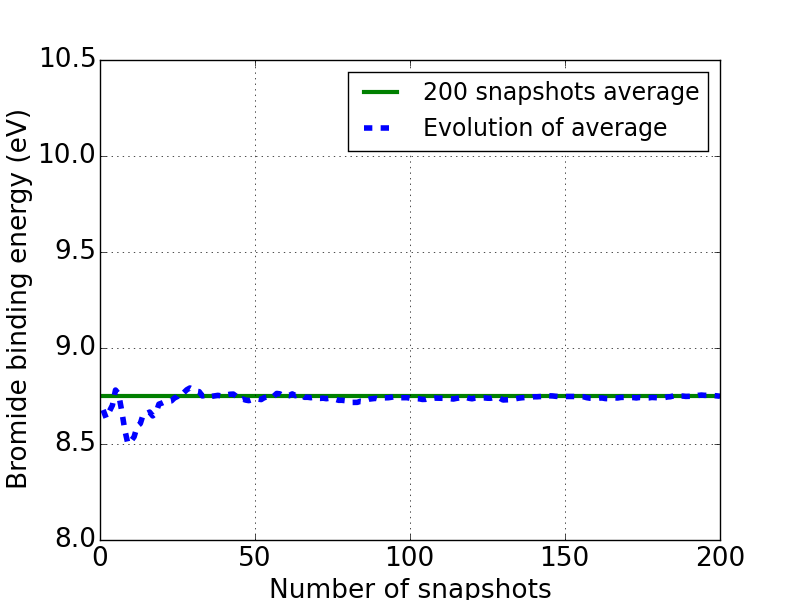}
       \caption{\ce{Br-}}\label{Fig:Br-BEevolution}
   \end{subfigure}
   \begin{subfigure}[b]{0.48\textwidth} 
       \centering \includegraphics[width=\textwidth]{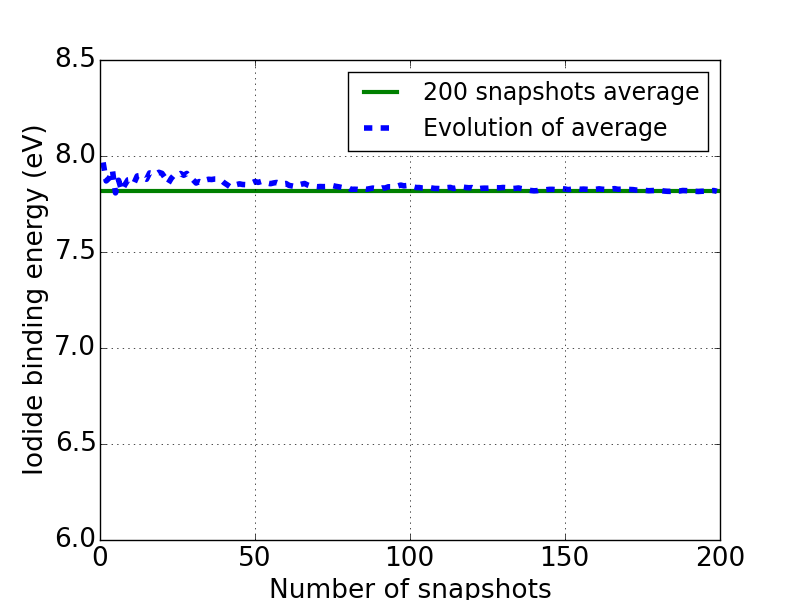}
       \caption{\ce{I-}}\label{Fig:I-BEevolution}
   \end{subfigure}

   \begin{subfigure}[b]{0.48\textwidth} 
       \centering \includegraphics[width=\textwidth]{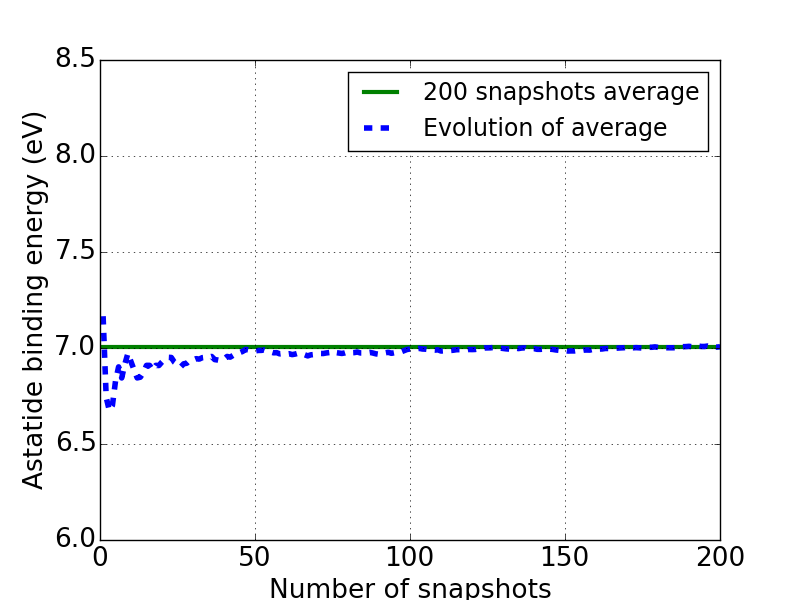}
       \caption{\ce{At-}}\label{Fig:At-BEevolution}
   \end{subfigure}
\caption{Evolution of the average halide binding energies computed at the DC SAOP level for the [\ce{X-}@\ce{(H2O)_{50}}] FnT models, across the 200 selected snapshots.~\cite{paper:figures}}
\label{Fig:X-BEevolution}
\end{figure}

The dynamics of halide droplets with 50 water molecules were obtained from classical molecular dynamics (CMD) simulations performed at T=\SI{300}{\kelvin} with runs of \SI{5}{\ns}, with the POLARIS(MD) code~\cite{Masella:2006da,Masella:2011dy,Masella:2013iw,Coles:2015er} using the polarized force fields previously developed for the halides~\cite{Real:2016bhb}, and were found to reproduce the radial distributions obtained from periodic calculations. We extracted 200~snapshots sampled each \SI{10}{\ps} the last \SI{1}{\ns} of the CMD trajectories. The autocorrelation functions of the halide binding energies shown in Figure~\ref{Fig:X-BEautocorr} die away instantaneously indicating that the 200 snapshots are temporarily uncorrelated, which justifies our present sampling choices. For all halides, the statistical BEs averages are converged with 100 snapshots only as seen from Figure~\ref{Fig:X-BEevolution}. To preserve a statistically uncorrelated sampling, the statistical ensemble comprises 100~snapshots with a twice as long sampling interval of \SI{20}{\ps}.

\section{SAOP DFT-in-DFT results for [\ce{X-}@\ce{(H2O)_50}] FnT models}
\begin{figure}[ht]
  \centering
   \begin{subfigure}[b]{0.48\textwidth} 
       \centering \includegraphics[width=\textwidth]{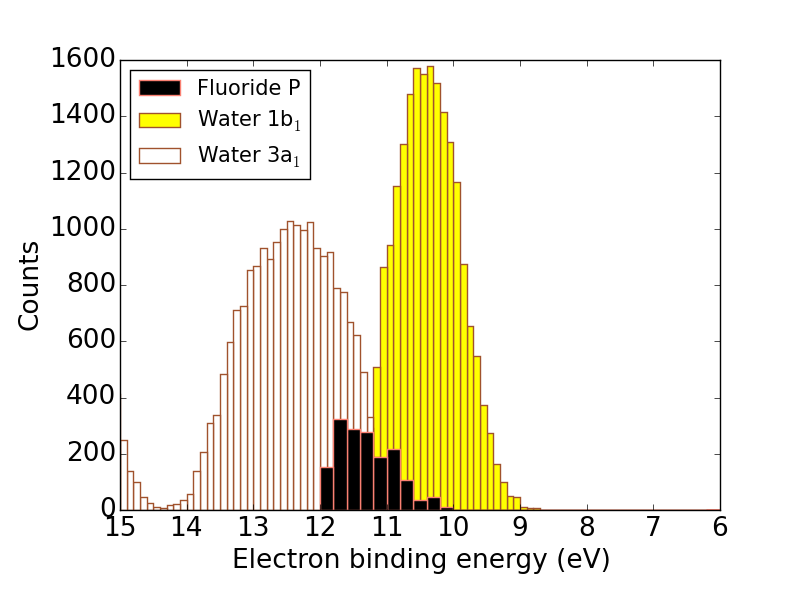}
       \caption{\ce{F-}}\label{Fig:F-SR-SAOP}
   \end{subfigure}
   \begin{subfigure}[b]{0.48\textwidth}
       \centering \includegraphics[width=\textwidth]{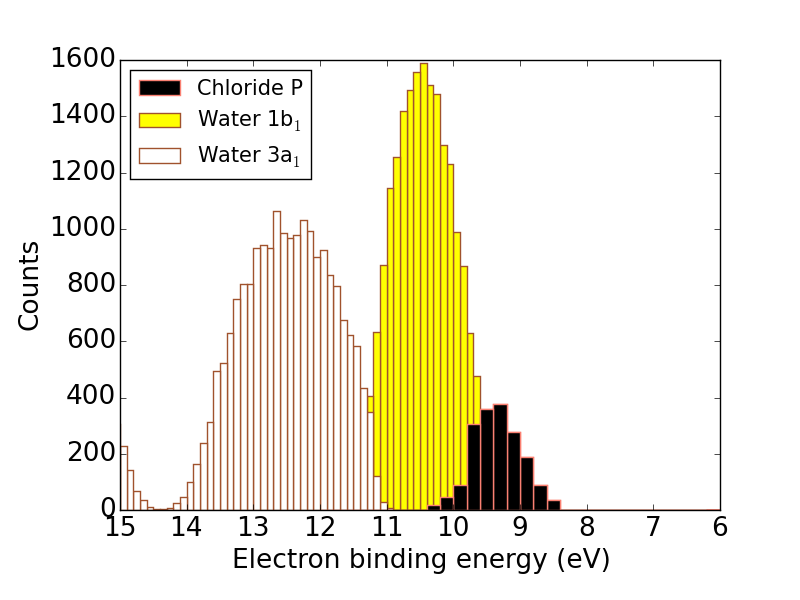}
       \caption{\ce{Cl-}}\label{Fig:Cl-SR-SAOP}
   \end{subfigure}

   \begin{subfigure}[b]{0.48\textwidth} 
       \centering \includegraphics[width=\textwidth]{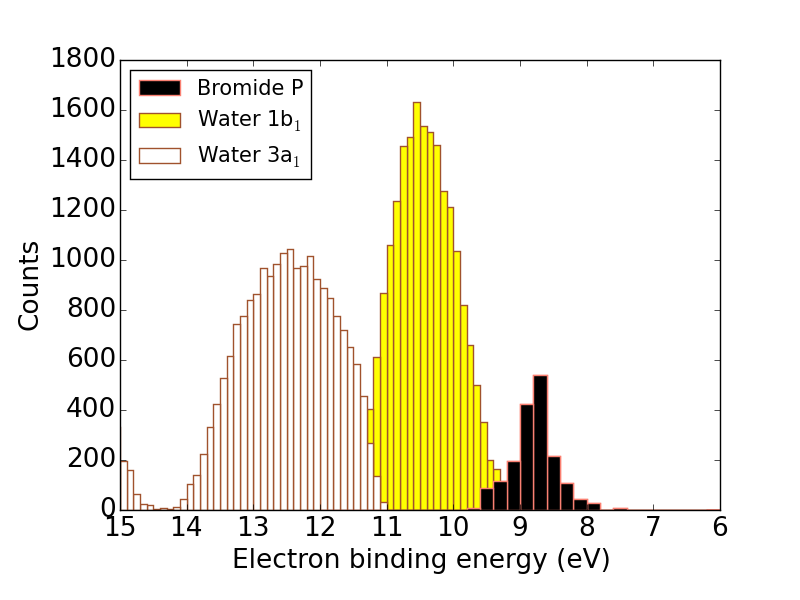}
       \caption{\ce{Br-}}\label{Fig:Br-SR-SAOP}
   \end{subfigure}
   \begin{subfigure}[b]{0.48\textwidth}
       \centering \includegraphics[width=\textwidth]{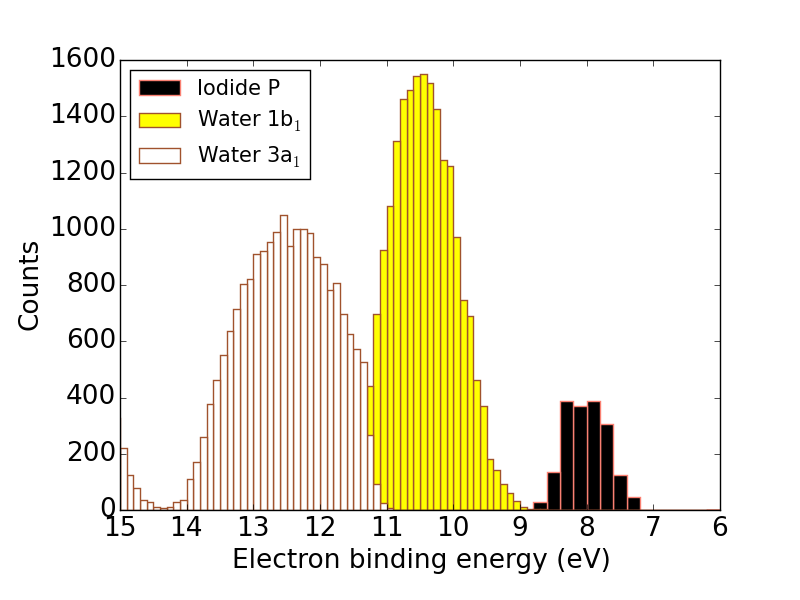}
       \caption{\ce{I-}}\label{Fig:I-SR-SAOP}
   \end{subfigure}

   \begin{subfigure}[b]{0.48\textwidth} 
       \centering \includegraphics[width=\textwidth]{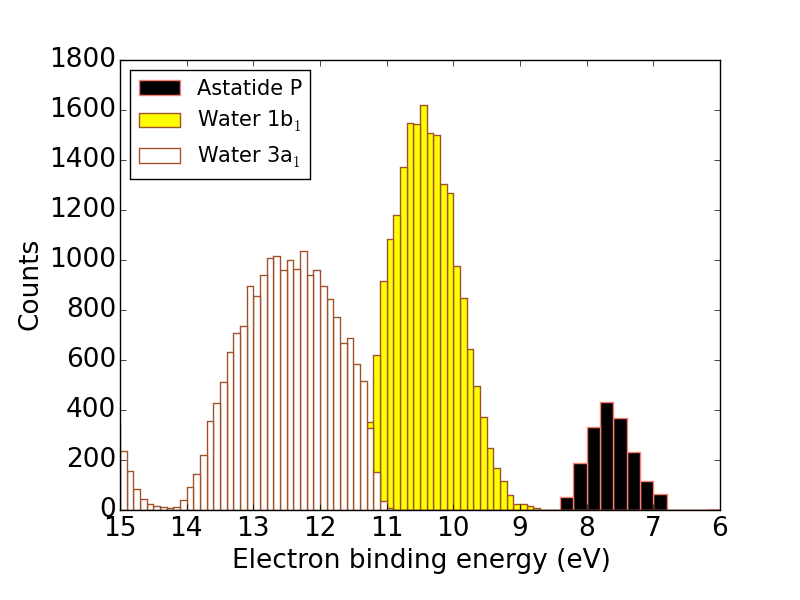}
       \caption{\ce{At-}}\label{Fig:At-SR-SAOP}
   \end{subfigure}
   
\caption{Electron binding energies spectra in \si{\eV} for the [\ce{X-}@\ce{(H2O)_{50}}] FnT systems over 200 snapshots, obtained from ZORA SAOP scalar relativistic calculations.~\cite{paper:figures}}
\label{Fig:X-SR-SAOP}
\end{figure}

\begin{figure}[ht]
  \centering
   \begin{subfigure}[b]{0.48\textwidth} 
       \centering \includegraphics[width=\textwidth]{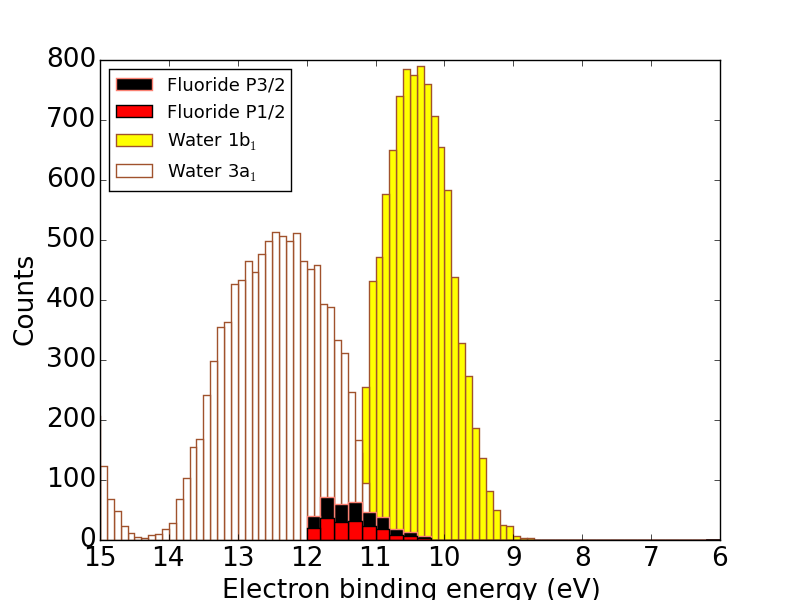}
       \caption{\ce{F-}}\label{Fig:F-SO-SAOP}
   \end{subfigure}
   \begin{subfigure}[b]{0.48\textwidth}
       \centering \includegraphics[width=\textwidth]{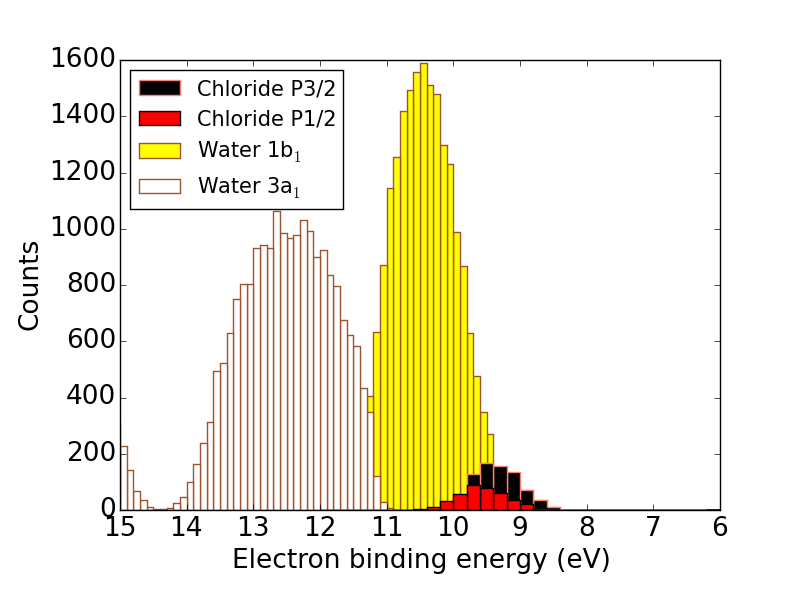}
       \caption{\ce{Cl-}}\label{Fig:Cl-SO-SAOP}
   \end{subfigure}

   \begin{subfigure}[b]{0.48\textwidth} 
       \centering \includegraphics[width=\textwidth]{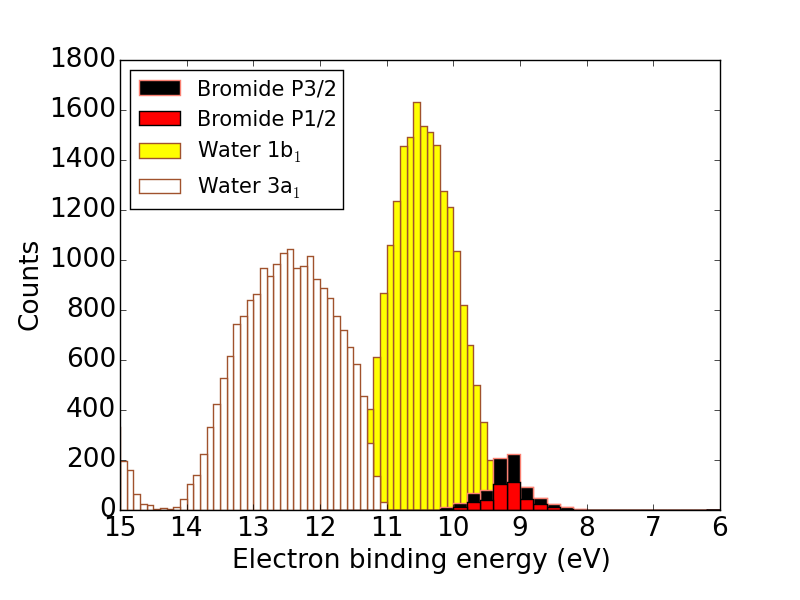}
       \caption{\ce{Br-}}\label{Fig:Br-SO-SAOP}
   \end{subfigure}
   \begin{subfigure}[b]{0.48\textwidth}
       \centering \includegraphics[width=\textwidth]{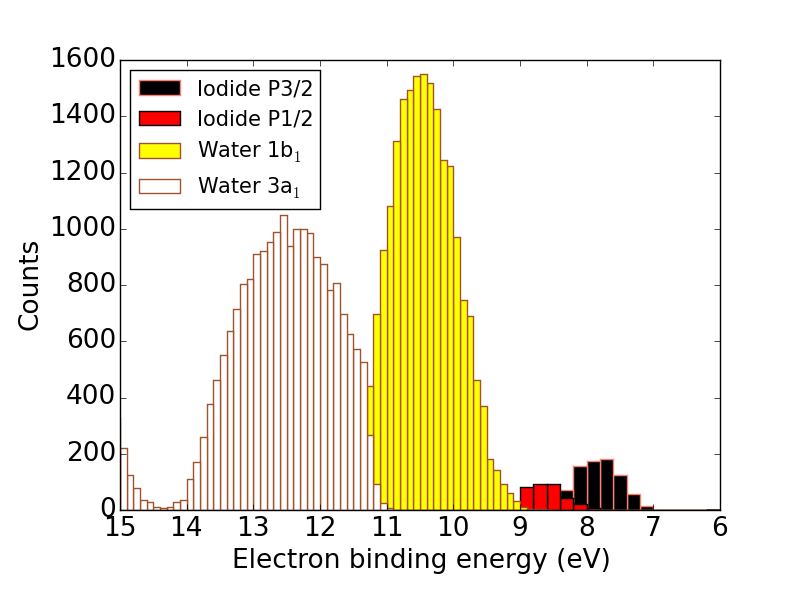}
       \caption{\ce{I-}}\label{Fig:I-SO-SAOP}
   \end{subfigure}

   \begin{subfigure}[b]{0.48\textwidth} 
       \centering \includegraphics[width=\textwidth]{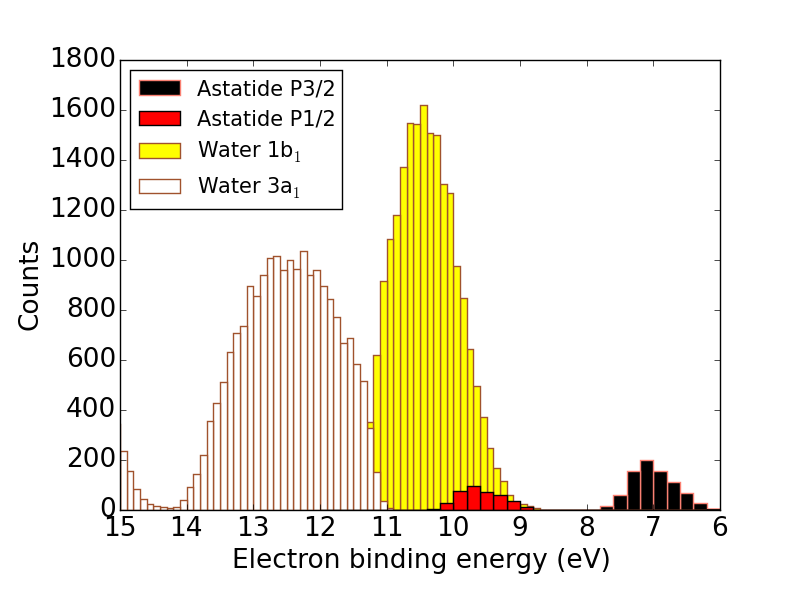}
       \caption{\ce{At-}}\label{Fig:At-SO-SAOP}
   \end{subfigure}
   
\caption{Electron binding energies spectra in \si{\eV} for the [\ce{X-}@\ce{(H2O)_{50}}] FnT systems over 200 snapshots, obtained from DC SAOP calculations with triple-zeta basis sets.~\cite{paper:figures}}
\label{Fig:X-SO-SAOP}
\end{figure}

To optimize the computational cost of the accurate DC-EOM-IP-CC calculations, it is relevant to explore the effect of spin-orbit coupling on the binding energies at the DFT-in-DFT level. Figures~\ref{Fig:X-SR-SAOP} and~\ref{Fig:X-SO-SAOP} report the BEs obtained at the scalar relativistic and spin-orbit levels, respectively. Spin-orbit coupling (SOC) has no effect on the water BE bands, but it increasingly separates the halogen $^2P_{1/2}$ and $^2P_{3/2}$ bands. In astatide, SOC amounts to about~\SI{2.5}{\eV}, making the $^2P_{1/2}$ peak overlap the water 1b$_{1}$ band.
 
\section{DC-EOM-IP-CC results, basis set effects}

\begin{figure}[ht]
\centering

   \begin{subfigure}[b]{0.48\textwidth} 
       \centering \includegraphics[width=\textwidth]{Valence_photoemission_spectrum_F-_EOMIP_WFT-in-DFT_TZ_6-15eV_0_50.png}
       \caption{\ce{F-} with triple-$\zeta$}\label{Fig:F-EOM-TZ}
   \end{subfigure}
   \begin{subfigure}[b]{0.48\textwidth} 
       \centering \includegraphics[width=\textwidth]{Valence_photoemission_spectrum_Cl-_EOMIP_WFT-in-DFT_TZ_6-15eV_0_50.png}
       \caption{\ce{Cl-} with triple-$\zeta$}\label{Fig:Cl-EOM-TZ}
   \end{subfigure}

   \begin{subfigure}[b]{0.48\textwidth} 
       \centering \includegraphics[width=\textwidth]{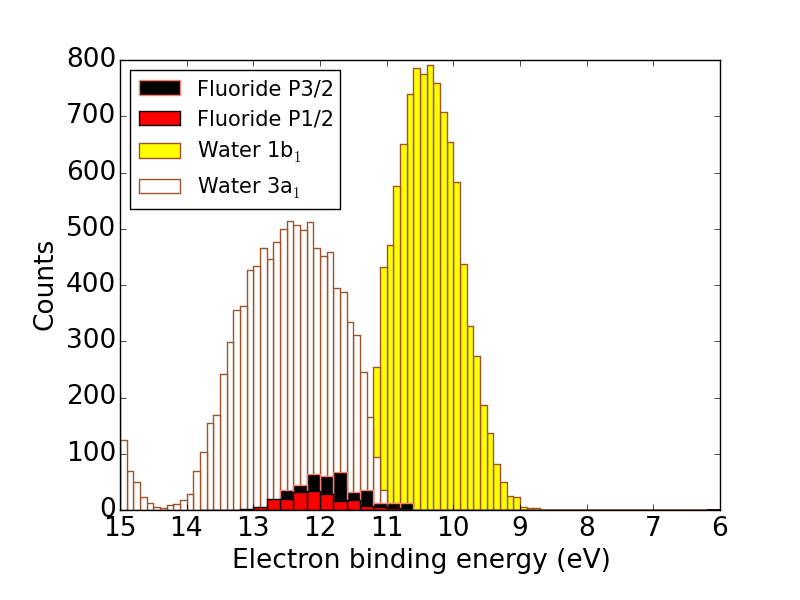}
       \caption{\ce{F-} with quadruple-$\zeta$}\label{Fig:F-EOM-QZ}
   \end{subfigure}
   \begin{subfigure}[b]{0.48\textwidth} 
       \centering \includegraphics[width=\textwidth]{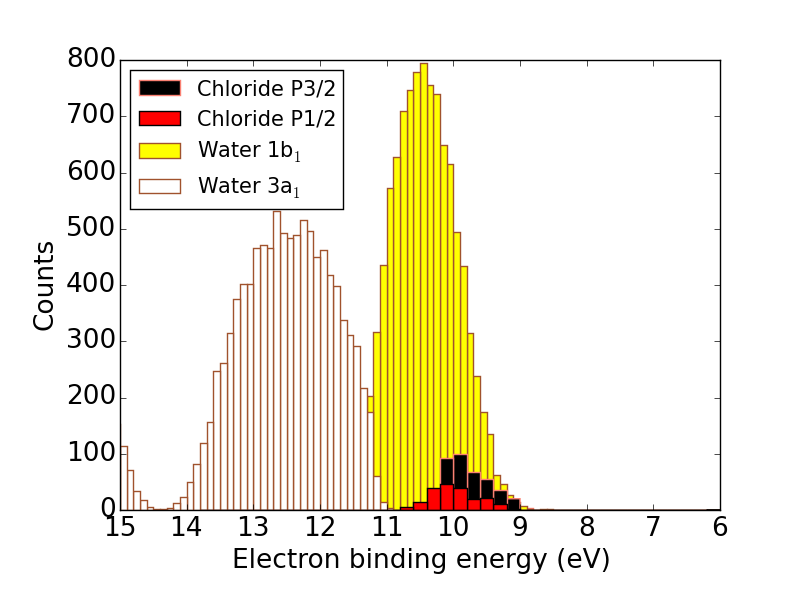}
       \caption{\ce{Cl-} with quadruple-$\zeta$}\label{Fig:Cl-EOM-QZ}
   \end{subfigure}

   \begin{subfigure}[b]{0.3\textwidth} 
       \centering \includegraphics[width=\textwidth]{Valence_photoemission_spectrum_Br-_EOMIP_WFT-in-DFT_TZ_6-15eV_0_50.png}
       \caption{\ce{Br-} with triple-$\zeta$}\label{Fig:Br-EOM-TZ}
   \end{subfigure}
   \begin{subfigure}[b]{0.3\textwidth} 
       \centering \includegraphics[width=\textwidth]{Valence_photoemission_spectrum_I-_EOMIP_WFT-in-DFT_TZ_6-15eV_0_50.png}
       \caption{\ce{I-} with triple-$\zeta$}\label{Fig:I-EOM-TZ}
   \end{subfigure}
   \begin{subfigure}[b]{0.3\textwidth} 
       \centering \includegraphics[width=\textwidth]{Valence_photoemission_spectrum_At-_EOMIP_WFT-in-DFT_TZ_6-15eV_0_50.png}
       \caption{\ce{At-} with triple-$\zeta$}\label{Fig:At-EOM-TZ}
   \end{subfigure}

\caption{Electron binding energies spectra in \si{\eV} for the [\ce{X-}@\ce{(H2O)_{50}}] systems over 100 snapshots. Comparison of halides BEs obtained from DC EOM-IP-CCSD with triple-zeta basis sets, and quadruple-zeta basis sets for \ce{F-} and \ce{Cl-}.~\cite{paper:figures}}
\label{Fig:X-EOM}
\end{figure}

\begin{figure}[ht]
\centering
\includegraphics[width=\textwidth]{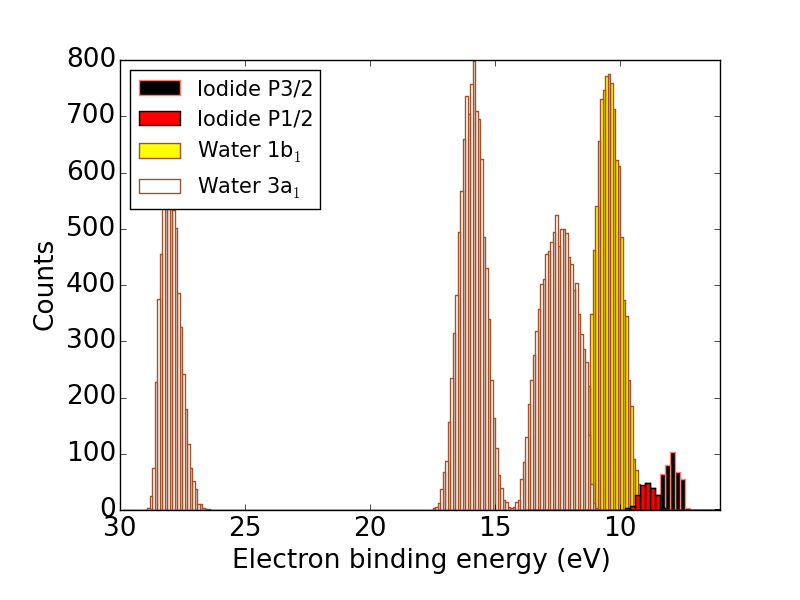}
\caption{Extended electron binding energies spectra in \si{\eV} for the [\ce{I-}@\ce{(H2O)_{50}}] with triple-zeta basis sets.~\cite{paper:figures}}
\label{Fig:I-whole-spectrum}
\end{figure}

\begin{figure}[ht]
\centering

   \begin{subfigure}[b]{0.48\textwidth} 
       \centering \includegraphics[width=\textwidth]{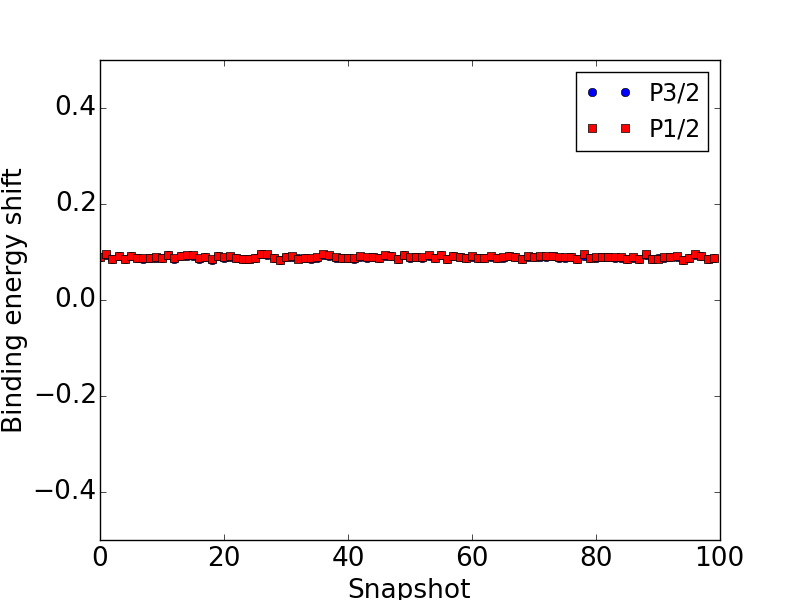}
       \caption{\ce{F-}}\label{Fig:F-EOM-TZ-QZ-shift}
   \end{subfigure}
   \begin{subfigure}[b]{0.48\textwidth} 
       \centering \includegraphics[width=\textwidth]{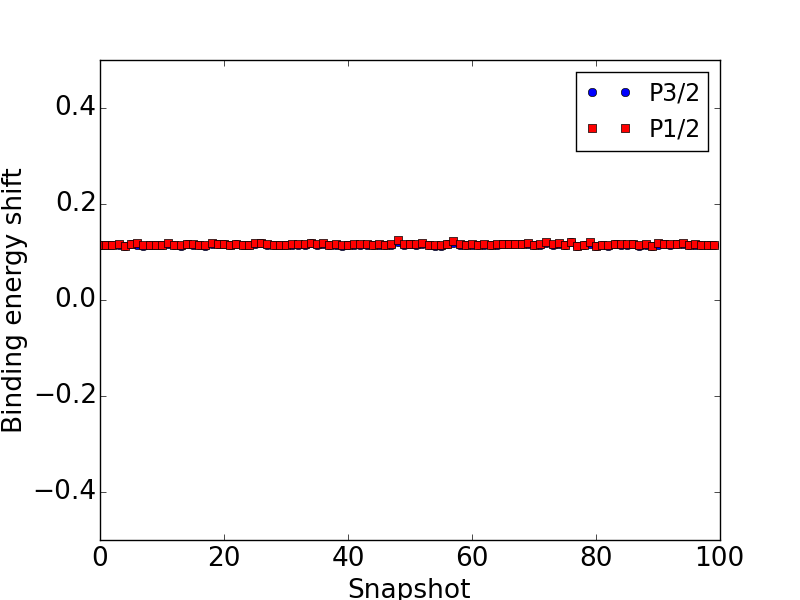}
       \caption{\ce{Cl-}}\label{Fig:Cl-EOM-TZ-QZ-shift}
   \end{subfigure}

\caption{Triple-zeta to quadruple-zeta shifts of the halide's DC EOM-IP-CCSD binding energies in \si{\eV}, for 100 snapshots of [\ce{F-}@\ce{(H2O)_{50}}] and [\ce{Cl-}@\ce{(H2O)_{50}}].}
\label{Fig:X-EOM-TZ-QZ-shift}
\end{figure}

Figure~\ref{Fig:X-EOM} shows the electron BEs spectra obtained with triple-zeta and quadruple-zeta basis sets for \ce{F-} and \ce{Cl-}. The increase of the basis set quality on the halide yields a slide increase of about \SI{0.1}{\eV} of the halide BEs. This energy shift is constant across the 100~snapshots as illustrated by Figure~\ref{Fig:X-EOM-TZ-QZ-shift}, and is of the same order of magnitude of that observed for bare halides, thus justifying that complete basis set (CBS) extrapolated values can be estimated from the sole atomic calculations reported in Table~\ref{Tab:IP-atomic-CBS}. Figure~\ref{Fig:I-whole-spectrum} represents the extended PES for the [\ce{I-}@\ce{(H2O)_{50}}] with triple-zeta basis sets in which water inner bands are shown.

\clearpage

\providecommand{\latin}[1]{#1}
\makeatletter
\providecommand{\doi}
  {\begingroup\let\do\@makeother\dospecials
  \catcode`\{=1 \catcode`\}=2 \doi@aux}
\providecommand{\doi@aux}[1]{\endgroup\texttt{#1}}
\makeatother
\providecommand*\mcitethebibliography{\thebibliography}
\csname @ifundefined\endcsname{endmcitethebibliography}
  {\let\endmcitethebibliography\endthebibliography}{}